\documentclass[11pt,a4paper]{article}
\usepackage[utf8]{inputenc}
\usepackage{newunicodechar}
\newunicodechar{β}{$\beta$}
\usepackage[total={155mm,262mm},top=17.5mm,left=27.5mm]{geometry}
\usepackage{graphicx}
\usepackage{mathtools}
 \usepackage{comment} 

\usepackage[backend=biber,style=numeric,sorting=none]{biblatex}
\usepackage{makecell}
\usepackage{marvosym}
\usepackage{amsmath}

\usepackage[normalem]{ulem}
\usepackage{comment}
\usepackage[dvipsnames]{xcolor}
\usepackage{subcaption}
\usepackage{pdfpages}
\usepackage{amsmath,amsfonts,amssymb}
\usepackage{booktabs}
\usepackage{colortbl}
\usepackage{hyperref}
\usepackage{times}
\usepackage{changepage}
\usepackage[capitalise]{cleveref}
\usepackage{siunitx}
\usepackage{authblk}
\usepackage{csquotes}
\usepackage{soul}
\usepackage{booktabs} 
\DeclareSIUnit\cells{\text{cells}}
\usepackage{multirow}
\usepackage{threeparttable}
\usepackage[utf8]{inputenc} 

\definecolor{myblue}{RGB}{0,100,200}
\definecolor{myred}{RGB}{204,102,0}
\definecolor{mygreen}{RGB}{0,200,50}
\hypersetup{
	colorlinks,
	linkcolor=myred,
	citecolor=myblue,
	urlcolor=myblue
}

\colorlet{tablerowcolor}{gray!10} 
\title{A Data-Enhanced Agent-Based Model for Simulating 3D Cancer Spheroid Growth: Integrating Metabolism and Mechanics
}
\author[1]{Pedro Garcia-Gomez}

\author[1]{Paula Guerrero-Lopez}
\author[2]{Silvia Hervas-Raluy}
\author[1*]{Jose Manuel Garcia-Aznar}

\affil[1]{Multiscale in Mechanical and Biological Engineering (M2BE), Dpt. of Mechanical Engineering, Aragon Institute of Engineering Research (I3A), University of Zaragoza, Zaragoza, Spain}

\affil[2]{Institute for Computational Mechanics, Technical University of Munich, Garching b. München, Germany}

\affil[*]{Corresponding author: silvia.hervas@tum.de}

\begin{document}

\maketitle
\section*{Abstract}
Cancer research has shifted from a purely gene-centric view to a more holistic understanding that recognizes the critical role of the tumour microenvironment, where mechanics and metabolism are key drivers of disease progression. However, the intricate interplay between these multifactorial mechanisms remains poorly understood. To address this gap, we present an agent-based computational model (ABM) that integrates tumour metabolism and mechanics to study 3D cancer spheroid growth. Our approach unifies the metabolism and mechanical aspects of tumour development within an integral model for cancer spheroid formation and growth. In addition to that, we performed a computational calibration of the parameters and tested the model versatility to reproduce different cellular behaviours. Our model reproduced qualitatively and quantitatively the experimental results of spheroid growth  obtained in the lab and also allowed to discern different dynamics that cancer cells can present under the same conditions, providing insight into the potential factors contributing to the variability in the size of spheroids. Furthermore, it also showed its adaptability to reproduce diferent cell lines and behaviours by tuning its parameters. This study highlights the significant potential and versatility of integrative modelling approaches in the field of cancer research, not only as a tool to complement in vitro studies, but also as independent tools to derive conclusions from the physical reality.

\textbf{Keywords:} Agent-based model; 3D tumor growth; Cancer metabolism; Cancer mechanics; Bayesian calibration

\section{Introduction}
Throughout the 20th century up until recent years, cancer research has primarily focused on either improving the therapies used for its treatment \cite{devita_two_2012}, or investigating the genetic aspect of the disease \cite{knudson_cancer_2002, boccia_systematic_2010}. Both approaches have been driven by the concept of cancer as a cellular condition rather than multi-faceted disease affected not only by problems in the replication of the genetic material of the cell, but by many different aspects that contribute to the expansion and development of the disease, such as the mechanic forces that act on the cells or the availability of nutrients. \cite{nia_physical_2020,bell_principles_2020}.  Recent studies have highlighted the critical role played by other factors such as cellular metabolism  and the tumour microenvironment \cite{martinez-outschoorn_cancer_nodate}, specifically, matrix stiffness and tumour cell metabolism \cite{sohrabi_microenvironmental_2023}. Therefore, a deeper understanding of how these multifactional mechanisms could be related would enhance the efficacy of existing therapies and treatments.\\

Cells obtain the energy needed for their physiological functions through a series chemical reactions encompassed in what is known as energy metabolism. These reactions produce adenosine tri-phospate (ATP), the main energy source of the cell. The mode of ATP production varies depending on the cell type and the micro-environmental conditions. Under oxygen-rich conditions, cells undergo aerobic metabolism and generate substantial amounts of ATP via oxidative phosphorylation (OXPHOS). In contrast, when oxygen availability is limited, oxidative phosphorylation cannot proceed, and ATP is produced predominantly through glycolysis in an anaerobic regime. Although less efficient in terms of energy yield, anaerobic metabolism is faster than its aerobic counterpart. Regular healthy cells tend to rely on the oxydative phosporilation  as their main way to produce energy, as it is the most efficient set of reactions. However, cancer cells have been observed to switch to the anaerobic regime even when there is oxygen available in the microenvironment. This phenomena is know as the Walburg effect \cite{zheng_energy_2012} \cite{jose_choosing_2011} .  The precise reasons underlying this metabolic reprogramming remain uncertain. Initially, it was hypothesized that this effect occurs due to a permanent malfunctioning of the oxidative phosphorylation \cite{isidoro_alteration_2004}. However, more recent theories suggest that this metabolic shift may confer a selective advantage to cancer cells \cite{gatenby_why_2004}. For instance, the anaerobic pathway produce lactate which acidifies the surrounding microenvironment. While normal cells are susceptible to damage and death under acidic conditions, cancer cells appear resilient, potentially giving them an advantage for proliferation\cite{lopez_lazaro_warburg_2008}.\\

The metabolism is far from being the only aspect apart from genetics that can affect cancer cells. The microenvironment in which cells reside also plays a pivotal role in regulating cancer development, both directly and indirectly.  Among its various components, the physical and mechanical properties of the extracellular matrix (ECM) are particularly influential. Features such as interstitial fluid pressure and tissue architecture have been shown to critically affect tumour growth, cellular behaviour, and metastatic potential \cite{nia_physical_2020}. These mechanical cues modulate not only the accessibility of nutrients and signaling molecules but also the ability of cancer cells to migrate, proliferate, and adapt to their surroundings, underscoring the complex interplay between the microenvironment and tumour progression \cite{plou_individual_2018}. One critical mechanical property of the ECM is its density, which can modulate various aspects of cancer cell physiology. High-density ECMs are associated with reduced cell motility, thereby impeding metastatic dissemination, which depends on the ability of cancer cells to migrate through the surrounding tissue \cite{zaman_migration_2006, wolf_physical_2013}. Moreover, ECM density affects the morphology and compactness of tumour spheroids \cite{hernandez-hatibi_quantitative_2025}. Denser matrices tend to promote more compact spheroid formation, which in turn limits the diffusion of essential nutrients and oxygen into the spheroid core, potentially resulting in hypoxic conditions and altered cellular behaviour \cite{kiran_mathematical_2009, vinci_advances_2012, }. These biophysical constraints underscore the integral role of the ECM in shaping cancer cell dynamics beyond purely genetic factors.\\

In recent years, computational models have emerged as a valuable tool for investigating the biological and physical mechanisms underlying various biological processes. In the context of cancer computational modelling, there are different approaches that normally can be used, distinguishing  continuous models \cite{hervas-raluy_tumour_2023, santagiuliana_simulation_2016, sciume_multiphase_2013} and cell-based or agent-based models \cite{ metzcar_review_2019,stephan_agent-based_2024}.  Continuous models tend to be faster and more computationally efficient, as they do not simulate individual cells but the homogenised properties of the tissue formed by the cells. Agent-based models, meanwhile, are usually more computationally expensive but they offer a more direct translation to biological observations. In these types of models, single cells are being simulated and their behaviour can be programmed and controlled individually. This allows single-cell differentiation and more spatial heterogeneity than continuous models.\\

In silico models offer notable advantages in terms of speed and flexibility, allowing for rapid and systematic variation of experimental conditions that may be difficult or time-consuming to reproduce using in vitro methods. However, computational tools are not yet sufficient as stand-alone research methods. To enhance their reliability and relevance, they are most effective when used in combination with in vitro or in vivo experiment data, which help to feed the  model parameters and validate simulation outcomes.\\

Computational studies focused exclusively on a single regulatory mechanism of cancer are inherently limited, as they fail to account for the collective and interconnected nature of the factors governing tumour growth. In this work, we present an agent-based computational model that integrates both tumour metabolism and mechanics, through the consideration of the regulatory mechanisms due to ECM mechanics, nutrients and oxygen availability. Incorporating these two effects within a unified framework presents several challenges, most notably the presence of numerous unknown parameters that require careful calibration with reliable experimental data  for validating model predictions. To address this, we performed in vitro experiments using 3D microfluidic devices to obtain tumour spheroids growth, providing quantitative data for calibrating the model. To estimate the unknown model parameters from experimental data, we employed a Bayesian inference strategy \cite{hervas-raluy_tumour_2023, merino-casallo_integration_2018}. This non-deterministic approach systematically explores the parameter space and compares model predictions with experimental observations to infer the most probable parameter values. Given the high computational cost associated with simulating the agent-based model, we utilized a surrogate modelling approach to enable efficient exploration of the parameter landscape. Specifically, we implemented a Gaussian process model as the surrogate model, allowing for an accurate approximation of the model output with significantly reduced computational demand.\\

The integration of in vitro and in silico approaches offers a powerful framework for studying complex biological systems such as tumour progression. In silico models enable hypotheses testing and  exploration of different conditions and meanwhile in vitro experiments provide physiologically relevant data. By combining both strategies, it is possible to not only improve model accuracy through data-enhanced calibration, but also to gain deeper insights  and underlying mechanisms present in the $\textit{in vitro}$ experiments.\\

The presentation of this manuscript is organized as follows. In Section \ref{sec:mat_and_meths}, we first present the experimental methodology carried out to asses spheroid growth patterns and provide the data necessary for the model calibration (Section \ref{sec:experiments}). We then provide a detailed description of the mechanical (Section \ref{sec:mech_model}) and metabolic (Section \ref{sec:met_model}) components of the model, as well as the statistical framework employed for calibration, including model order reduction, sensitivity analysis and Bayesian infernece techniques (Section \ref{sec:stat_methods}). The results and the predictive capabilities of the model, as well as the exploration of new scenarios, are presented and discussed in Section \ref{sec:results} . Finally, in Section \ref{sec:conclusions}, we summarize the key conclussions of this study and propose potential directions for future research.

\section{Materials and Methods}\label{sec:mat_and_meths}

\subsection{Experimental setup}\label{sec:experiments}
We conducted a series of in vitro experiments using the A549 cell line, which is derived from a human lung carcinoma. Individual A549 cells were cultured in 3D microfluidic devices embedded in collagen gels of varying densities to investigate their growth behaviour and spheroid formation. Images of the microfluidic devices were captured daily for seven days following cell seeding. These images were subsequently analysed to quantify the size of the resulting spheroids. \\

\subsubsection{Microfluidic devices fabrication}

The microfluidic devices used in this study are fabricated from polydimehtyl-siloxane (PDMS), a biocompatible and transparent material, following the protocol described by Shin et al \cite{shin_microfluidic_2012}. The design includes a central chamber, where a collagen hydrogel mimicking the tissue matrix and containing the cells is confined, and two lateral channels through which the nutrients are introduced, as shown in Figure \ref{fig:exp_fig_1} a). The geometry is patterned onto a silicon wafer using an SU-8 master mould and then replicated in PDMS. The PDMS was prepared by mixing the base and curing agent at a 10:1 weight ratio, followed by curing at 80°C overnight. Once cured, the devices were trimmed and punched to create access ports for the channels. Finally, the PDMS structures were bonded to the glass bottoms of 35 mm Petri dishes via plasma treatment, and subsequently coated with poly-D-lysine (PDL) to enhance collagen matrix adhesion within the device.

\subsubsection{Hydrogel preparation and cell seeding}
\begin{figure}[h]
    \centering
    \includegraphics[width=0.9\textwidth]{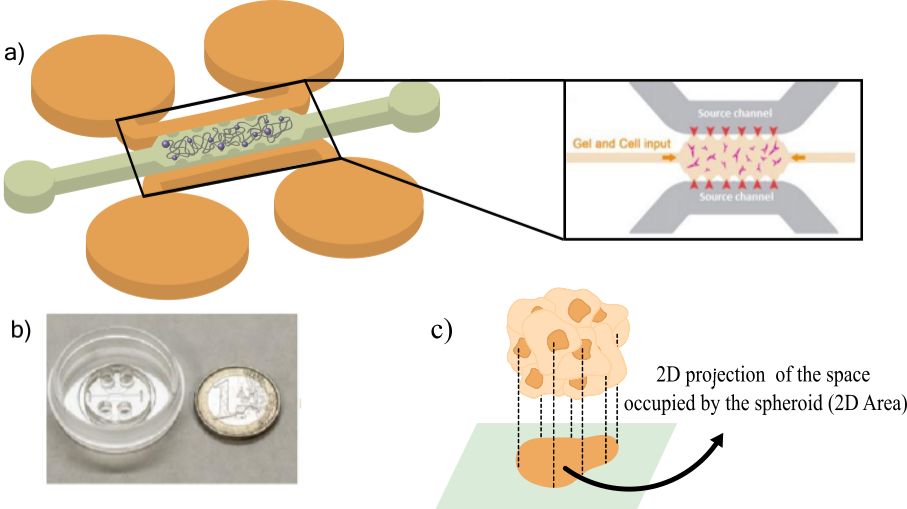}
    \caption{a) Geometry and design of the microfluidic devices described in the present work, with a central chamber containing the gel with the 3D cells embedded and two side channels to feed the nutrients. \cite{plou_individual_2018} b) Size comparison between a microfluidic chip and a 1 euro coin. c) A scheme of the 2D area projection of the spheroid measured in silico and in vitro.}
   \label{fig:exp_fig_1}
\end{figure}

The lung adenocarcinoma cell line A549 was cultured with Dulbecco's modified Eagle's medium at 4.5 g/L glucose and supplemented with 10$\%$ fetal bovine serum and 1$\%$ of penicillin/streptomicin. Cells were incubated at 37 °C with 5$\%$ CO$_2$ until 80$\%$ confluence was reached for use in the experiment. When optimal confluence was reached, cells were trypsinised, centrifuged (1200 rpm, 5 minutes) and passed through a 40 µm cell strainer in order to ensure removal of cell aggregates. Subsequently, cells were counted using a Neubauer chamber and added to the collagen mix to leave a final concentration of 0.2 x 106 cell/mL. A type I collagen-based matrix was used for the 3D cell culture, using the protocol by Shin et al. \cite{shin_microfluidic_2012} The hydrogel consisted of a mixture at 4 ºC of 10X DPBS, collagen at a final concentration of 6 mg/mL, 0.5M NaOH to adjust pH to 7.5 and the cells.  This mix was introduced into the central chamber, as shown in Figure \ref{fig:exp_fig_1} a), and left to polymerize at 37 ºC in humid boxes, turning the device every 5 minutes for at least 20 minutes. Finally, the lateral channels were hydrated with culture medium. These devices where conserved in a CO$_2$ incubator at 37 ºC for 9 days and the media of the two lateral chambers was changed every 2 days.

\subsection{Image acquisition and data treatment}\label{sec:image_acq}
The spheroid growth was monitored with a Leica DM IL Led microscope. Daily pictures were taken of the central chamber at 4X magnification in brightfield. Later, these images were processed with ImageJ \cite{schneider_nih_2012} and the spheroids 2D projected area was segmented with the in-house semi-automatic Segmentation3D App developed by C. Borau using Matlab (Mathworks, Natick, CA, US) as described by Alamán-Díez  et al \cite{alaman-diez_collagen-laponite_2023}. The data obtained were processed and represented using GraphPad Prism 8. A scheme of this measure is presented in Figure \ref{fig:exp_fig_1} c).

\subsection{Computational approach.} \label{sec:models}
The computational model is divided  in two coupled models: A mechanic  and a metabolic model. 
\subsubsection{Mechanical model: Modelling cell-cell and cell-ECM interactions.}\label{sec:mech_model}

The mechanical model used in this study builds upon the framework proposed by Gonçalves et al.\cite{goncalves_extracellular_2021} which extends the built-in agent-based mechanical model of PhysiCell to incorporate the cell-matrix interactions. The ECM plays a key role in regulating multiple cellular processes. In the context of tumour spheroid development and growth, the density of the  ECM can influence both the size and growth rate of the spheroid.  Additionally, the ECM can also affect how far individual cells migrate \cite{plou_individual_2018}, thereby affecting spheroid compactness, as increased migratory capacity often results in less aggregated spheroid structures.\\

\begin{figure}[h]
    \centering
    \includegraphics[width=1\textwidth]{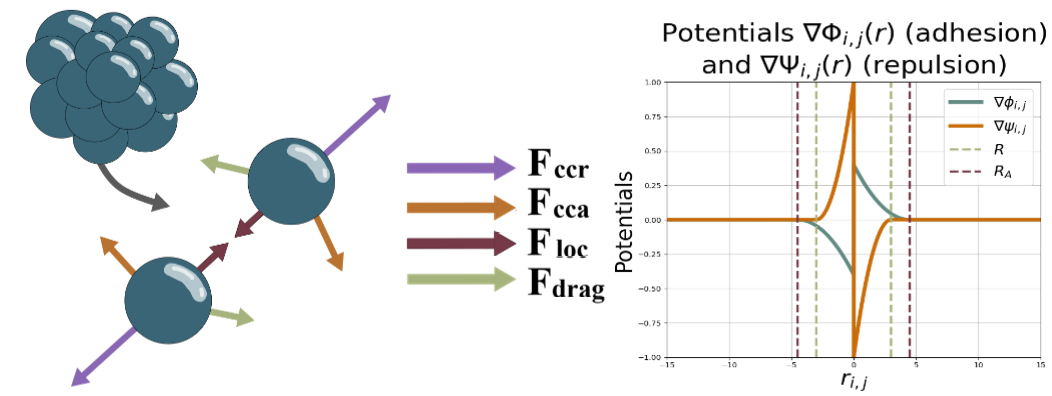}
    \caption{a) Scheme of the forces that act in the cell as part of the mechanic model. b) Graphic representation of the attraction and repulsion potential between two cells where the x-axis represents r = $x_j$ - $x_i$, being $x_i$ and $x_j$ the position of the particle $i$ and $j$ . $R$ an $R_A$ are the characteristic distances of the two potentials. These two distances define the range of action of both potential functions.}
    \label{fig:forces}
\end{figure}

In the computational model, each cell is represented as a fixed-size agent whose position evolves over time based on the net balance of  mechanical forces acting upon it at each time step.These forces arise from two types of interactions: intercellular forces and interactions between the cells and the ECM (Figure~\ref{fig:forces}.a). The interaction between cells is governed by a balance of two opposing forces: a cell-to-cell adhesion force and a cell-to-cell repulsion force. These forces are derived from corresponding potential functions, as illustrated in Figure~\ref{fig:forces}.b. The interaction between cells and the ECM is modeled by other two main forces: a random locomotive force, which drives cell movement, and a drag force, which resists motion (Figure~\ref{fig:forces}). The mechanical properties of the ECM are incorporated into the model via the drag force, which is defined as proportional to the collagen viscosity \cite{valero_combined_2018}. In this framework, the ECM is treated as a continuous viscous medium that produces mechanical resistance, without explicitly modelling individual collagen fibers. \cite{goncalves_extracellular_2021}\\

The model considered in this study is non-inertial, as the inertial term can be neglected. This assumption is justified by the fact that the timescale over which forces reach equilibrium is significantly shorter than other characteristic cellular timescales, such as those associated with cell division or cell death \cite{ghaffarizadeh_physicell_2018}. Under this assumption, the equilibrium equation for each can be written as: 

\begin{equation}
    \sum_{j\in N(i)} (\textbf{F}^{ij}_{cca} + \textbf{F}^{ij}_{ccr} +\textbf{F}^{i}_{loc} +\textbf{F}^{i}_{drag})\approxeq 0, 
    \label{eq:equilibrium_og}
\end{equation}
being $N$ the total number of cells, $i$ and $j$ indexes that go through all the cells $i, j \in [1,N]$. $\mathbf{F}_{cca}$ and $\mathbf{F}_{ccr}$ are the adhesion and repulsive cell-to-cell forces respectively, $\mathbf{F}_{loc}$  is the random locomotive force and $\mathbf{F}_{drag}$ is the drag force of the extracellular matrix. The drag and locomotive forces act in the whole domain of the simulation. However, the two cell-cell interaction forces only act when the cells are at certain distances from each other. These two forces are given by the following expressions. For the adhesion interaction we formulate:

\begin{equation}
    \mathbf{F}_{cca}^{ij} = - C_{cca} \, \nabla \Phi\!\left( r = (\mathbf{x}_j - \mathbf{x}_i) \right),
\end{equation}

and for the repulsion interaction:

\begin{equation}
    \mathbf{F}^{ij}_{ccr} = - C_{ccr} \nabla \Psi\!\big(r = (\mathbf{x}_j - \mathbf{x}_i)\big),
\end{equation}

 where $C_{cca}$ and $C_{ccr}$ are the adhesion and repulsion coefficients that control the magnitude of the two forces and $\mathbf{x_i}$ and $\mathbf{x_j}$ are the positions of the particles $i$ and $j$. $\Phi$ and $\Psi$ are two potentials that depend on the distance between the cells and are given by the following expressions:
\begin{equation}
     \nabla \Phi_{i,j}(\textbf{r}) =
     \begin{cases}
     \left( 1- \frac{|\textbf{r}|}{R_A}  \right)^{2} \frac{\textbf{r}}{|\textbf{r}|}, & \text{if } |r| \leq R_A, \\
     0, & \text{otherwise.}
     \end{cases}
\end{equation}

\begin{equation}
     \nabla \Psi_{i,j}(\textbf{r}) =
     \begin{cases}
     -\left( 1- \frac{|\textbf{r}|}{R_d}  \right)^{2} \frac{\textbf{r}}{|\textbf{r}|}, & \text{if } |\textbf{r}| \leq R_d, \\
     0, & \text{otherwise.}
     \end{cases}
\end{equation}

Both potentials are described in the right panel of the Figure \ref{fig:forces}. $R_d$ is defined as the sum of the radius of the two particles that are interacting ($R_d = r_i + r_j$), while $R_A$ is defined as $R$ plus the maximum adhesion distance ($\delta$) that is established depending on the potential range of action ($R_A = R+\delta$). The values of the constants here  are detailed in Table \ref{table:physical_params}. \\

A drag force has been implemented, to model the interaction between the cells and the ECM, as the density of collagen of the ECM reduces the velocity of movement of the cells. The drag force is described as the Stokes' law: 

\begin{equation}
    \mathbf{F}^{(i)}_{\text{drag}} = -6 \pi R \eta \mathbf{v}_i
    \label{eq:drag}
\end{equation}

where $\mathbf{v_i}$ is the velocity of the $i$ cell, R is the radius of the cells and $\eta$ represents the dynamic viscosity of the collagen gel, which depends on the density of the collagen fibers of the gel and was characterised by Valero et al. in \cite{valero_combined_2018}.\\

The last force that needs to be considered is the locomotive force ($\mathbf{F_{loc}}$), which has been characterised using the experiments done by Plou et al.\cite{plou_individual_2018}. These experiments correspond to single cell migration tests where the cells are separated sufficiently for not interacting with each other. In this particular context, equation \ref{eq:equilibrium_og} can be reduced as the interaction terms can be neglected:\\

\begin{equation}
    \sum_{j\in N(i)} (\textbf{F}^{i}_{loc} +\textbf{F}^{i}_{drag})\approxeq 0
    \label{eq:equilibrium}
\end{equation}

 From this, and using the equation \ref{eq:drag} we can obtain the expression of the velocity for a single cell only interaction with the matrix: 

\begin{equation}
    \mathbf{v}_i = \frac{\mathbf{F}^{i}_{\text{loc}}}{-6 \pi R \eta}
    \label{eq:force}
\end{equation}

The $\textbf{F}^{i}_{loc}$ was considered a polinomical function and its parameters were calibrated by \cite{goncalves_extracellular_2021} using the experiments from \cite{plou_individual_2018}, as was cited previously. The polinomical function obtained was:

\begin{equation}
    F_{loc}(p)=1.56 p^3 + 3.27p^2 + 0.07p+0.06
    \label{eq:loc}
\end{equation}

where $p$ is a random numbre from 0 to 1.\\

Finally, starting from equation \ref{eq:equilibrium_og}, and applying Stokes' law along with the computation of all forces described in this section, the velocity of each cell can be determined at every simulation step according to the following expression: 

\begin{equation}
    \mathbf{v}_i = \frac{1}{6 \pi R \eta} 
    \sum_{j \in N(i)} \left( 
        \mathbf{F}^{ij}_{\mathrm{cca}} + 
        \mathbf{F}^{ij}_{\mathrm{ccr}} +
        \mathbf{F}^{i}_{\mathrm{loc}}
    \right)
    \label{eq:vel}
\end{equation}

\begin{equation}
    \mathbf{v}_i=\frac{1}{6\Pi R \eta}\sum_{j\in N(i)} (\mathbf{F}^{ij}_{cca} + \mathbf{F}^{ij}_{ccr} +(1.56 p^3 + 3.27p^2 + 0.07p+0.06))
    \label{eq:vel}
\end{equation}

\begin{table}[h]
\centering
\begin{tabular}{c|c|c|c|c}
\hline
\textbf{Symbol} & \textbf{Parameter} & \textbf{Value} & \textbf{Units} & \textbf{Reference} \\ \hline\hline

$C_{cca}$ & Adhesion coefficient  & 7.2   & -     & \cite{ghaffarizadeh_physicell_2018} \\ \hline
$C_{ccr}$ & Repulsion coefficient & 380   & -     & \cite{ghaffarizadeh_physicell_2018} \\ \hline
$R$       & Cell radius           & 6     & $\left[\mu m\right]$ & \cite{ghaffarizadeh_physicell_2018} \\ \hline
$R_A$     & Adhesion radius       & $1.25R$ & $\left[\mu m\right]$ & \cite{ghaffarizadeh_physicell_2018} \\ \hline
$\eta$    & Dynamic viscosity     & [7.96, 18.42, 39.15] & $\left[\mathrm{Pa \cdot s}\right]$ & \cite{valero_combined_2018} \\ \hline

\end{tabular}
\caption{List of physical parameters used in the model.}
\label{table:physical_params}
\end{table}

\subsubsection{Metabolical model: An ATP-driven model.}\label{sec:met_model}
\begin{figure}[h]
    \centering
    \includegraphics[width=0.75\textwidth]{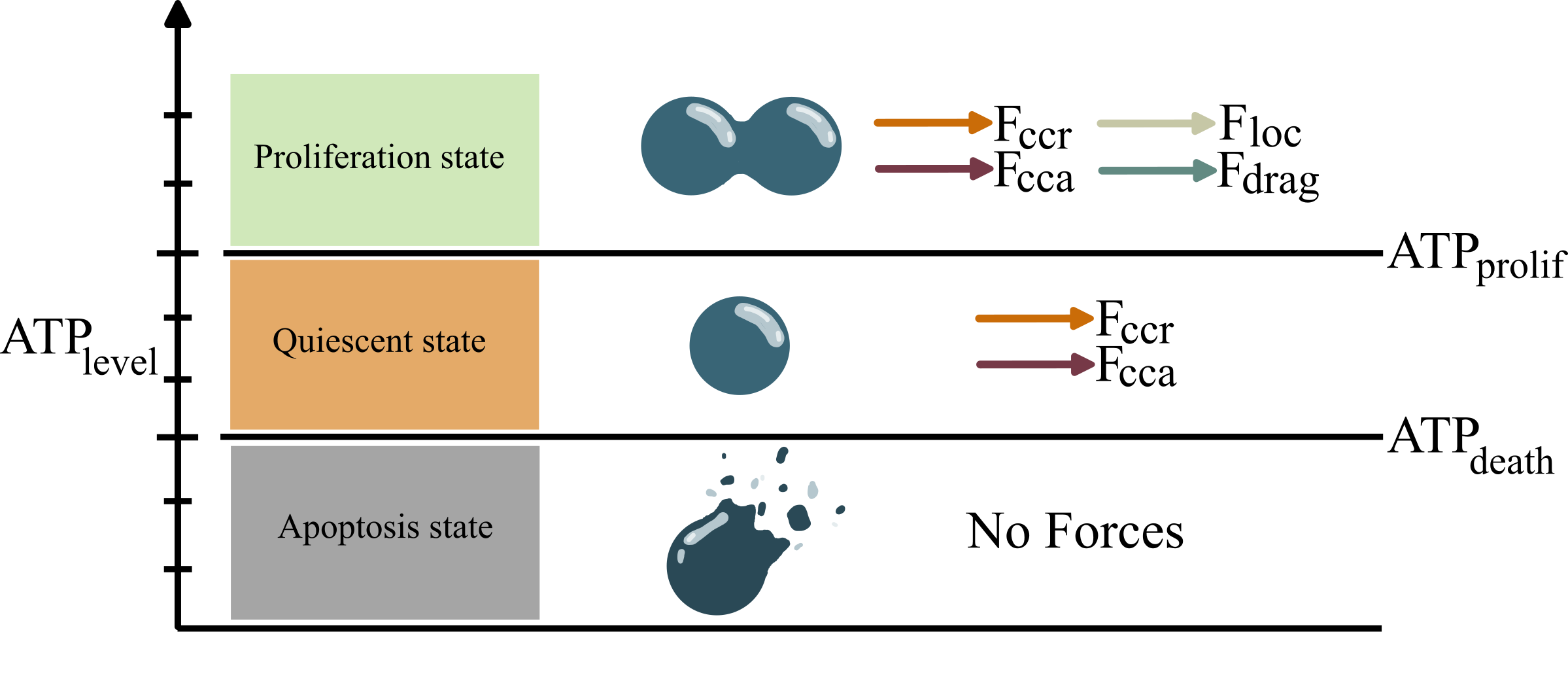}
    \caption{Scheme of the different states the cell can have in the model. $ATP_{prolif}$ level establish over which level of ATP the cells proliferate and migrate. $ATP_{death}$ establish below which level cells die. If the cell's ATP levels are between these two levels, cells are quiescent and they do not migrate nor proliferate. Depending on the ATP level of the cells, different forces play a rol. }
    \label{fig:levels}
\end{figure}

ATP, the primary energy source for most cellular processes,  also functions as a regulator of key activities such as cell division, metabolism, movement, and apoptosis \cite{eguchi_intracellular_nodate}, \cite{zamaraeva_cells_2005}, \cite{larsson_importance_2000}. Given ATP's central role in cellular metabolism, our metabolic model is structured around this molecule. At each time step, ATP levels are quantified for every individual cell and compared against two predefined thresholds. These levels determine whether the cells are apoptotic, quiescent or proliferating. Therefore, interactive mechanical forces can vary depending on the cell state as it is indicated in Figure \ref{fig:levels}.\\

The molecular and chemical pathways through which cells produce ATP can vary depending on cell type and the specific phase of the cell cycle.  However, most cells derive their ATP from glucose via two major processes: glycolysis and oxidative phosphorylation. Beyond these primary pathways, a complex network of additional reactions provides essential substrates for ATP synthesis. Therefore, we propose a simplified model, based on the one introduced by Jacquet et al. \cite{jacquet_reduced_2023}, which focuses on the main reactions responsible for ATP production.\\

The energy metabolism of a human cell can be summarized in five main processes, each one involving multiple coupled reactions.  The abbreviation used in the chemical equations that summarized these processes are shown in Table \ref{table:chemistry}, and the five process mentioned are described below:

\begin{table}[h]
\centering
\begin{tabular}{c|c}
\hline
\textbf{Abbreviation}     & \textbf{Molecule}                           \\ \hline\hline
\textit{Glu}              & Glucose                                     \\ \hline
\textit{$NAD^+$} & Nicotinamide adenine dinucleotide (oxidized form) \\ \hline
\textit{NADH}             & Nicotinamide adenine dinucleotide (reduced form)  \\ \hline
\textit{ADP}              & Adenosine diphosphate                       \\ \hline
\textit{ATP}              & Adenosine triphosphate                      \\ \hline
\textit{Pyr}              & Pyruvate                                    \\ \hline
\textit{$P_i$}             & Phosphate molecule                          \\ \hline
\textit{$H_2O$}            & Water                                       \\ \hline
\textit{CoA-SH}           & Coenzyme A                                  \\ \hline
\textit{Acetyl-CoA}       & Acetyl coenzyme A                           \\ \hline
\textit{H+}       & Hydrogen ion                          \\ \hline
\textit{FAD}       & Flavin adenin dinucleotid                          \\ \hline\textit{FADH}       & Reduced flavin adenin dinucleotid                          \\ \hline\textit{GDP}       & Guanosin diphosphate                    \\ \hline\textit{GTP}       & Guanosin triphosphate                     \\ \hline\textit{Q}       & Ubiquinone                       \\ \hline\textit{$QH_{2}$}       & Ubiquinol (reduced ubiquinone)                       \\ \hline\textit{$CO_{2}$}       & Carbon dioxide                         \\ \hline
\textit{Lac}           & Lactate                                 \\ 
\hline
\end{tabular}
\caption{List of molecular abbreviations used in the model.}
\label{table:chemistry}
\end{table}

\begin{itemize}
    \item Glycolisis: in this process,  glucose is converted to pyruvate \cite{chandel_glycolysis_2021}. During this sequence of reactions, 2 molecules of ATP are produced. This anaerobic process occurs in the cytoplasm and can be summarized as follows:\\

    \begin{equation}
        \begin{multlined}
            Glu + 2 NAD^{+} + 2ADP +2 P_{i} 
             \leftrightarrow 2 Pyr + 2 NADH + 2 H^{+} + 2 ATP + 2 H_{2}O
        \end{multlined}
        \label{eq:gluc}
    \end{equation} 

    \item Lactic acid fermentation: This anaerobic reaction is described by the following equation:\\

    \begin{equation}
        \begin{multlined}
            Pyr + NADH \leftrightarrow
             Lac + NAD^{+}
        \end{multlined}
        \label{eq:Ana}
    \end{equation}

    Although this reaction does not generate ATP directly, it regenerates $NAD^+$, a critical substrate for glycolysis \cite{wang_lactate_2021}.\\
    
    \item Pyruvate oxidation:  The pyruvate oxidation transform the pyruvate produced in the glycolisis into acetyl-coenzime-A:

    \begin{equation}
            Pyr + NAD^{+} + CoA-SH \leftrightarrow
             Acetyl-CoA + NADH
        \label{eq:pyr-ox}
    \end{equation}

    \item Citric acid cycle: also known as the Krebs cycle (equation \ref{eq:krebs}), takes place in the mitocondria and transform the Acetil-coenzime-A produced during pyruvate oxidation into the necessary metabolites for oxidative phosphorilation.

    \begin{equation}
        \begin{multlined}
        \mathrm{Acetyl{-}CoA} + 3\,\mathrm{NAD}^{+} + \mathrm{FAD} 
           + \mathrm{GDP} + P_{i} + 2\,H_{2}O \\
        \longleftrightarrow \; 
        4\,\mathrm{NADH} + \mathrm{FADH}_{2} + 4\,H^{+} 
        + \mathrm{GTP} + 3\,CO_{2}
        \end{multlined}
        \label{eq:krebs}
    \end{equation}

    \item Oxydative phosphorilation: The oxidative phosphorilation takes place in the electron transport chain of the walls of the mitocondria and it uses the metabolites produced at the Krebs cycle (equation \ref{eq:krebs}) at the ATP syntase to produce ATP molecules from ADP: 

\begin{equation}
\begin{multlined}
5\mathrm{NADH} + \mathrm{QH_2} + 17\mathrm{ADP} + 17\mathrm{P_i}
+ 3\mathrm{O_2} + 5\mathrm{H^+} \\
\leftrightarrow
5\mathrm{NAD^+} + \mathrm{Q} + 6\mathrm{H_2O} + 17\mathrm{ATP}
\end{multlined}
\label{eq:OXPHOS}
\end{equation}

\end{itemize}

In order to save computational resources, the last three reaction (\ref{eq:pyr-ox}, \ref{eq:krebs} and \ref{eq:OXPHOS}) have been summarised into a single aerobic reaction of production of energy. This was done because we are just interested in the production of ATP, not the intermediate metabolites, and these three metabolic reactions take place in a chain where only the last one produces ATP molecules. The equation that sums up these three metabolic processes is:\\

\begin{equation}
    \begin{multlined}
        Pyr + NADH + 3 O_{2}\leftrightarrow
         17ATP + NAD^{+}
    \end{multlined}
    \label{eq:aerobic_reaction}
\end{equation}

To simulate computationally the consumption of energy done by the cells due to their different functions (migration, proliferation...), we will implement one last equation. The cellular consumption of ATP equation:\\

\begin{equation}
        ATP \rightarrow \emptyset
        \label{eq:ene}
\end{equation}

After all, the computational model here presented has 4 final equations (see Table \ref{table:met_eq}). These equations were implemented in COPASI as ordinary differential equations (ODEs).\\

\begin{table}[h]
\centering
\begin{tabular}{c|c}
\hline
\textbf{Metabolic reaction} & \textbf{Equation} \\ \hline\hline

Glycolysis & 
$\mathrm{Glu} + 2\,\mathrm{NAD}^{+} 
\;\longleftrightarrow\;
2\,\mathrm{Pyr} + 2\,\mathrm{NADH} + 2\,ATP$ \\ \hline

Aerobic ATP production & 
$\mathrm{Pyr} + \mathrm{NADH} + 3\,O_{2} 
\;\longleftrightarrow\;
17\,ATP + \mathrm{NAD}^{+}$ \\ \hline

Lactic acid fermentation & 
$\mathrm{Pyr} + \mathrm{NADH} 
\;\longleftrightarrow\;
\mathrm{Lac} + \mathrm{NAD}^{+}$ \\ \hline

Cellular ATP consumption & 
$ATP \;\longrightarrow\; \varnothing$ \\ \hline

\end{tabular}
\caption{Table of the metabolic equations implemented in our computational model. Obtained from equations \ref{eq:gluc}, \ref{eq:Ana}, \ref{eq:aerobic_reaction} and \ref{eq:ene} after some simplifications.}
\label{table:met_eq}
\end{table}

\subsubsection{Coupling of metabolical and mechanical models}
The two models explained in the subsections \ref{sec:mech_model} and \ref{sec:met_model} operate independently and are invoked at different parts of the code. However, they are coupled in order to introduce dependencies between them and thereby provide a more accurate representation of reality. The coupling strategy is represented in Figure \ref{fig:levels}. The metabolic model, through the equations of consumption and production of metabolites explained in \ref{sec:met_model} determines at each time step of the simulation the intracellular ATP levels of each individual cell. ATP levels in each cell are then compared against the thresholds described in Figure~\ref{fig:levels} to determine whether the cell state is dead, quiescent, or proliferating. As it can be seen in Figure~\ref{fig:levels}, depending on the ATP level, different cell functions are activated and correspondingly different forces are considered:

\begin{itemize}
	\item ATP levels below $ATP_{death}$ threeshold: The cells below this ATP level are in the apoptotic state. This state do not have any force associated, as dead cells do not interact with their environment.
	\item ATP levels between $ATP_{death}$ and $ATP_{prolif}$ threeshold: Cells in this ATP range are quiescent. The quiescent state has forces associated only with the interaction with other cells (cell-cell adhesion force ($\mathbf{F}_{cca}$) and cell-cell repulsion force ($\mathbf{F}_{ccr}$)), as quiescent cells do not migrate.
	\item  ATP levels above $ATP_{prolif}$ threeeshold: The cells which internal ATP is above this ATP level are proliferating. These cells have forces related with the interaction between cells and forces derived from the movement and the relation with the extracellular matrix (cell-cell adhesion force ($\mathbf{F}_{cca}$), cell-cell repulsion force ($\mathbf{F}_{ccr}$), lococotive force ($\mathbf{F}_{loc}$) and drag force ($\mathbf{F}_{drag}$).
\end{itemize}

The computational model described in this section contains numerous parameters whose exact values are not well established. This uncertainty arises from the difficulty of accurately measuring these parameters, their variability across different cell lines, and the limited availability of data in the literature. These parameters can be categorized into two main groups: metabolic and non-metabolic parameters (see Table \ref{table:params}). A brief description of these parameters is presented here:\\

\begin{itemize}
    \item \textbf{Metabolic parameters}. These include the rates at which the metabolic processes outlined in Section \ref{sec:met_model} occur within the cell: $k_{glu}$ for glycolysis (Eq. \ref{eq:gluc}), $k_{aer}$ for the Krebs cycle (Eq. \ref{eq:krebs}), and $k_{ana}$ for lactic acid fermentation (Eq.\ref{eq:Ana}), as well as the rate of cellular energy consumption, $k_{ene}$ (Eq. \ref{eq:ene}). These parameters are not directly measurable, and the estimations available in the literature lack the accuracy required to confidently assign specific values.

    \item \textbf{Non-metabolic parameters}. These include the ATP thresholds associated with cell death ($ATP_{death}$) and proliferation ($ATP_{prolif}$), as illustrated in Figure \ref{fig:levels}, as well as the cellular proliferation rate ($k_{prolif}$). The ATP thresholds are intrinsic to the model and cannot be directly measured. While estimations for the proliferation rate are available in the literature \cite{lu_biological_2009, chary_maximizing_2022}, this parameter is subject to variability due to experimental conditions and different measure methods, introducing a degree of uncertainty. Therefore, these parameters also need to be estimated.
    
\end{itemize}

\begin{table}[h]
\centering
\renewcommand{\arraystretch}{1.4}
\begin{tabular}{c|c|c|c}
\hline
\textbf{Category} & \textbf{Parameter} & \textbf{Description} & \textbf{Units} \\ \hline\hline

\multirow{4}{*}{Metabolic} 
 & $k_{glu}$  & Rate of glycolysis               & $\left[\tfrac{\mathrm{ml}^2}{\mathrm{mmol}^2 \cdot \mathrm{min}}\right]$ \\ \cline{2-4}
 & $k_{aer}$  & Rate of Krebs cycle              & $\left[\tfrac{\mathrm{ml}^2}{\mathrm{mmol}^2 \cdot \mathrm{min}}\right]$ \\ \cline{2-4}
 & $k_{ana}$  & Rate of lactic acid fermentation & $\left[\tfrac{\mathrm{ml}^4}{\mathrm{mmol}^4 \cdot \mathrm{min}}\right]$ \\ \cline{2-4}
 & $k_{ene}$  & Energy consumption rate          & $\left[\tfrac{\mathrm{ml}}{\mathrm{mmol}\cdot \mathrm{min}}\right]$ \\ \hline

\multirow{3}{*}{Non-metabolic} 
 & $k_{prolif}$ & Proliferation rate          & $\left[\tfrac{1}{\mathrm{min}}\right]$ \\ \cline{2-4}
 & $ATP_{prolif}$  & ATP proliferation threshold & dimensionless \\ \cline{2-4}
 & $ATP_{death}$  & ATP death threshold         & dimensionless \\ \hline

\end{tabular}
\caption{List and classification of the parameters of the model.}
\label{table:params}
\end{table}

\subsection{Model exploration via uncertainty quantification} \label{sec:stat_methods}
Accurate estimation of these parameters is crucial, as they can significantly influence the model's predictions. Given the absence of definitive values from the literature, we employ a Bayesian estimation framework, informed by experimental data, to derive probabilistic estimates for these parameters. In order to implement this Bayesian inference process more efficiently, we quantified the influence of the parameters in our model to identify those that could be fixed for the subsequent analysis . We also replace the high-fidelity model with a Gaussian process to reduce the computational cost of uncertainty quantification analyses.
\subsubsection{Sensitivity analysis} \label{subsec: parames_importance}

To assess the influence of each model parameter on the quantity of interest (the projected area of the spheroid), we employ a two-step sensitivity analysis. First, we perform a marginalization analysis in which each parameter is varied independently. This approach allows to observe the isolated effect of each parameter on model behaviour. Subsequently, we compute the Sobol sensitivity indices, which provide a quantitative measure of both the individual (first-order) and interactive (higher-order) contributions of each parameter to the output of the computational model. Together, these analyses enable us to identify the parameters that have the greatest impact on the selected quantity of interest and to disregard those with negligible influence, thereby refining the model and focusing subsequent analyses on the most relevant parameters.\\

Marginalization over a specific parameter entails varying this parameter in the range of possible values considered and check all the possible combinations of the rest of the $N-1$ parameters for the specific value of the studied parameter. Then, the mean of all these combinations of outcomes of this combination of parameters should be calculated to check the marginalized output of the model:\\
\begin{equation}
Pred. Mean_i = \frac{\sum^M_{j=0}{GP\{x_i,x_1^j,..,x_{i-1},x_{i+1},...,x_N^j\}}}{M}
\label{eq:marg}
\end{equation}

$M$ denotes the number of random combinations for the marginalization process, $N$ corresponds to the total number of parameters and $GP$ is  the surrogate model explained in the following subsection. \\

\subsubsection{Gaussian process} \label{subsec:gp}
Gaussian Processes (GPs) are powerful surrogate models that enable rapid estimation of a model’s output \cite{schulz_tutorial_2018-1}, eliminating the need to execute the full model each time a new set of parameters is evaluated. The primary goal of a Gaussian Process is, given a set of training data $D_{training} = \{t^1_i,t^2_i, ..., t^J_i, y_i\}_{i=1}^N$ where $\{t_i\}_{i=1}^J$ represents the input parameters, $y_i$ the corresponding output, $N$ the number of training samples, and $J$ the number of input parameters, to infer the probability distribution of the output at a new, unseen point $x_{\text{predict}}$ \cite{melkumyan_sparse_nodate}.\\

A GP is defined by its mean function and a covariance matrix ($\Gamma$).  The covariance matrix is constructed using the training data and has a square shape structure with dimensions $N \times N$, corresponding to the number of training points. This matrix is generated through a covariance function $\Gamma(t_i, t_j)=k(t_i, t_j)$, commonly referred as the kernel function.The choice of kernel function is crucial, as it determines how correlations between data points are modeled and directly influences the behaviour of the GP. The kernell function can take various forms, depending on the characteristics of the data we are trying to reproduce. In this work, the squared exponential covariance function (RBF: Radial Basis Function) was employed.

\begin{equation}
    k(t_i , t_j)=\sigma^2 \exp\left( -\sum_{i=1}^{M} \frac{(t_i - t_j)^2}{2l_i^2} \right)
    \label{eq:rbf}
\end{equation}
where $\sigma ^2$ denotes the variance and $l_i$ represents the characteristic length scale of each parameter. These quantities are the hyperparameters of the Gaussian Process. During training, the hyperparameters are optimized to maximize the likelihood of the model fitting the training data.\\

To assess the predictive performance of the GP metamodel, we use the Nash–Sutcliffe efficiency coefficient \cite{nash_river_1970}, defined as:\\

\begin{equation}
    Q^2=1-\frac{\sum(m_{GP}(\theta)-f(\theta))^2}{\sum(m_{GP}(\theta)-\overline{f})^2}
    \label{eq:nash}
\end{equation}
being $\overline{f}$:\\

\begin{equation}
   \overline{f}= \frac{1}{N_T}\sum f(\theta)
\end{equation}
where $N_T$ the number of testing samples. The Nash-Sutcliffe has a range from $-\infty$ to 1, where values closer to 1 indicate high predictive accuracy, while negative values suggest poor model performance \cite{nash_sutcliffe_2018}.\\

\subsubsection{Bayesian estimation}\label{subsec:bayesian}
Bayesian estimation was selected over other deterministic parameter estimation methods because it provides not only point estimates of the parameters but also their full probability distributions. Access to the full distribution is particularly important in biological systems, which are inherently stochastic. Reporting a single point estimate for a biological parameter could lead to overconfident and potentially misleading conclusions.

The Bayesian estimation framework is based on Bayes' theorem, which can be stated as:

\begin{equation}
    P(A|B) = \frac{P(B|A)·P(A)}{P(B)}=\frac{P(B|A)·P(A)}{\sum_A{P(B|A)·P(A)}}\propto P(B|A)·P(A)
    \label{eq:bayes}
\end{equation}

In the context of Bayesian estimation of parameters, the Bayes theorem can be reformulated as:

\begin{equation}
    p(mathbf{\theta}|\mathbf{y_{obs}}) \propto p(\mathbf{y_{obs}}|\mathbf{\theta})·p(\mathbf{\theta})
    \label{eq:probs}
\end{equation}

Here, $\mathbf{y_{obs}}$ represents the experimental data used for calibration and $\mathbf{\theta}$ are the parameters of our model. $p(\mathbf{\theta}|\mathbf{y_{obs}})$ is known as the posterior distribution and is the main objective of the Bayesian calibration process: it describes the probability distribution of the parameters conditioned on the observed data. $p(\mathbf{\theta})$ represents the prior distribution, which encodes the existing knowledge or assumptions about the parameters before observing the new data. The form of this prior distribution depends on the amount and quality of prior information available.\\

The term $p(\mathbf{y_{obs}}|\mathbf{\theta)}$ is the likelihood function \cite{van_de_schoot_bayesian_2021}, which quantifies the probability of observing the data given a set of parameter values. It links the experimental observations ($\mathbf{y}_{obs}$) to the computational model output. In this study, it is assumed that the experimental data follow a normal distribution centered around the model predictions $f(\mathbf{\theta})$:

\begin{equation}
    \mathbf{y}_{obs,i} \approx \mathcal{N} (f(\mathbf{\theta}), \sigma_N)
    \label{eq:distrt_normal}
\end{equation}

As each experimental data is assumed to be independently drawn from this normal distribution, the likelyhood can be written as the multiplication of the probability distribution of the individual observations as follows:\\

\begin{equation}
    p(y_{obs}|f(\theta))= \prod_{i=1}^{N} \mathcal{N}(y_{obs,i}|f(\theta), \sigma_N^2)=\prod_{i=1}^{N}\frac{1}{\sqrt{2\pi\sigma_N^2}}exp \left(-\frac{(y_{obs,i}-f(\theta))^2}{2\sigma_N^2}\right)
    \label{eq:likelyhood}
\end{equation}

To characterize the posterior distribution of the model parameters, multiple simulations are required to adequately sample the parameter space. The model here presented has a high computational cost. Due to this, a surrogate model  (GP) (Section \ref{subsec:gp}) is employed to approximate the model output $f(\mathbf{\theta})$ in the likelihood function (\ref{eq:likelyhood}).

\subsection{Computational implementation}\label{subsec:comptools}

For the implementation of the model, we have selected PhysiCell \cite{ghaffarizadeh_physicell_2018} as our main computational tool. PhysiCell is an open-source agent-based cell simulation program. This computational tool has already implemented many mechanical and phenotipic functionalities easily adaptable to the characteristics of the cancer cells we are simulating. Furthermore, PhysiCell offers seamless integration with other computational tools, which is crucial when developing coupled computational models like the one proposed in this work. PhysiCell has already demonstrated its value as a reliable and effective platform for cancer research, not only as a complement to experimental studies conducted in the lab, but also as a standalone modelling tool capable of delivering powerful insights. \cite{ponce-de-leon_optimizing_2022} \cite{ruscone_multiscale_2023}.\\

\begin{figure}[h]
    \centering
    \includegraphics[width=0.85\textwidth]{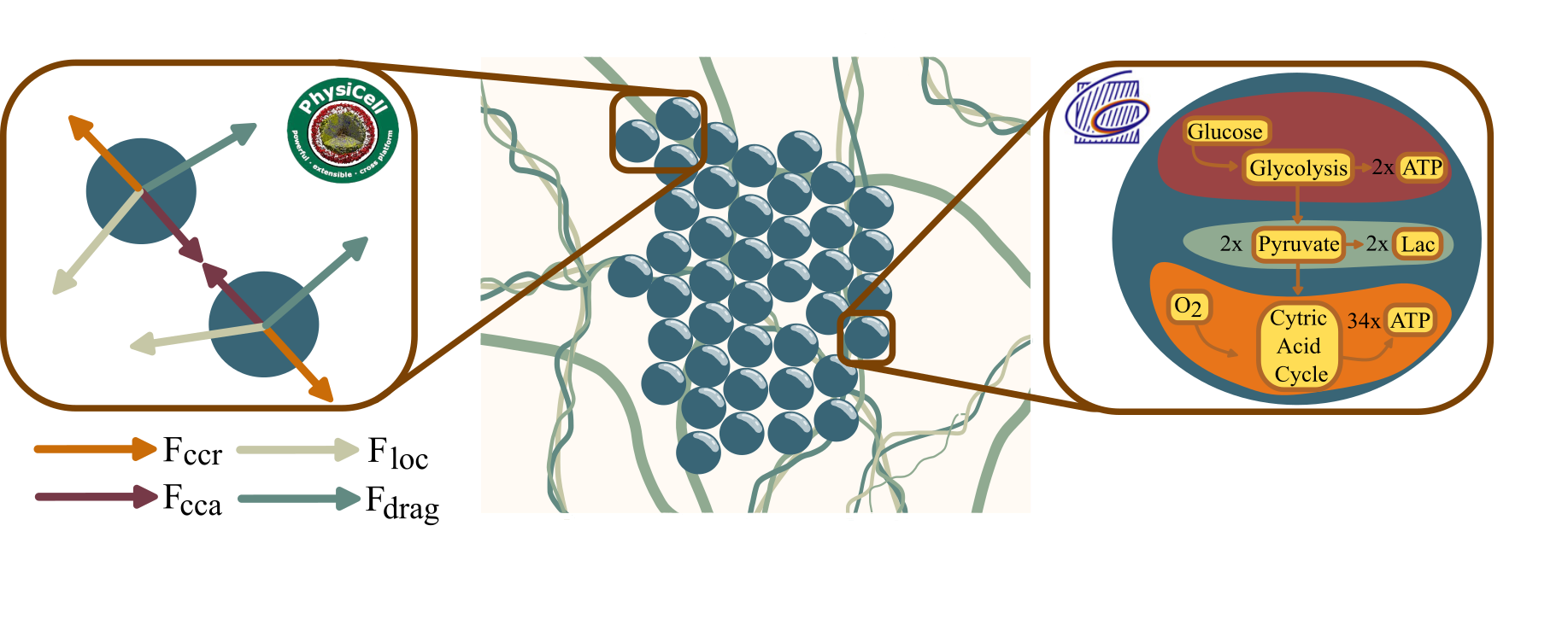}
    \caption{Schematic representation of the coupling of the mechanical and metabolical models.}
    \label{fig:coupling}
\end{figure}

The mechanical part of the model has been entirely implemented in PhysiCell, however, for the metabolic part, an external tool was integrated in the PhysiCell mechanical model. This external tool is Copasi \cite{hoops_copasicomplex_2006}, a software application built for the construction and simulation of biochemical networks and equations using SBML System Biology Markup Language), an open data format for the simulation of computational systems biology. A squematic representation of the integration, the model and the computational tools used is portrayed in Figure \ref{fig:coupling}.\\

\begin{figure}[h]
    \centering
    \includegraphics[width=0.9\textwidth]{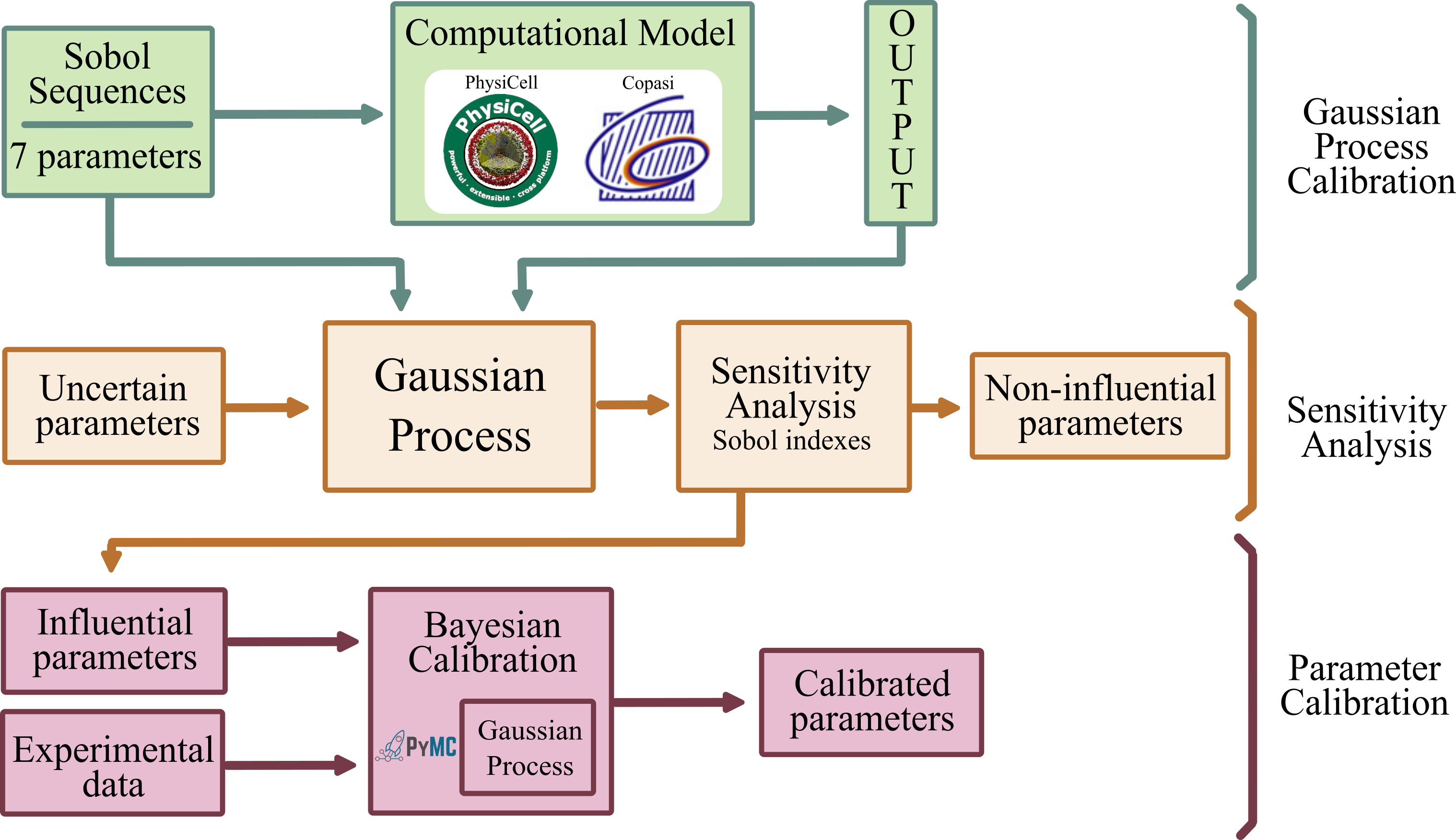}
    \caption{General scheme of the calibration process. First, a Gaussian process is trained using outputs from multiple simulations of the computational model with different parameter combinations (green boxes). This surrogate model is then employed to perform a sensitivity analysis, identifying influential and non-influential parameters (yellow boxes). Finally, Bayesian calibration is carried out using only the GP with the experimental data available to obtain the calibrated parameter set (red boxes).}
    \label{fig:bayes_esq}
\end{figure}

For the implementation of the sensitivity analysis and the calibration process described in Section~\ref{sec:mat_and_meths}, we employed the PyMC probabilistic programming library~\cite{abril-pla_pymc_2023}. The integration of the different models and statistical modules was carried out using PhysiCool~\cite{goncalves_physicool_2023}. An overview of the overall structure of our workflow—including the computational model, the sensitivity analysis, and the calibration process—is provided in Figure~\ref{fig:bayes_esq}.

\section{Results}\label{sec:results}
In this section, we first present the results of the microfluidic experiments conducted to study tumour spheroid growth within 3D microfluidic devices. We then proceed with a sensitivity analysis of the mathematical model to assess the relative influence of each parameter on the computational model's output. Subsequently, we describe the parameter estimation procedure used to calibrate the model against the experimental data. Finally, we evaluate the model’s ability to reproduce the experimental observations using the estimated parameter values.
\subsection{Experimental data of tumour spheroid growth}\label{subsec:experiments} 
Brightfield images of the tumour-on-chip device were collected at days 1, 4, 7, and 9 to track the temporal evolution of spheroid formation and progression (Figure \ref{fig:marg}a). Four independent replicas of the same experiment were performed to ensure reproducibility. Cells were seeded in the microfluidic devices using diffusors to favor single-cell deposition rather than pre-aggregated clusters. Once seeded, the devices were incubated at 37 °C with 5 \%  CO$_2$ for 9 days, and cells were supplied daily with glucose-rich medium (6 mg/ml) to support proliferation. Images were acquired every 24 hours with a Leica DM IL Led microscope.\\

\begin{figure}[h]
    \centering
    \includegraphics[width=1\textwidth]	{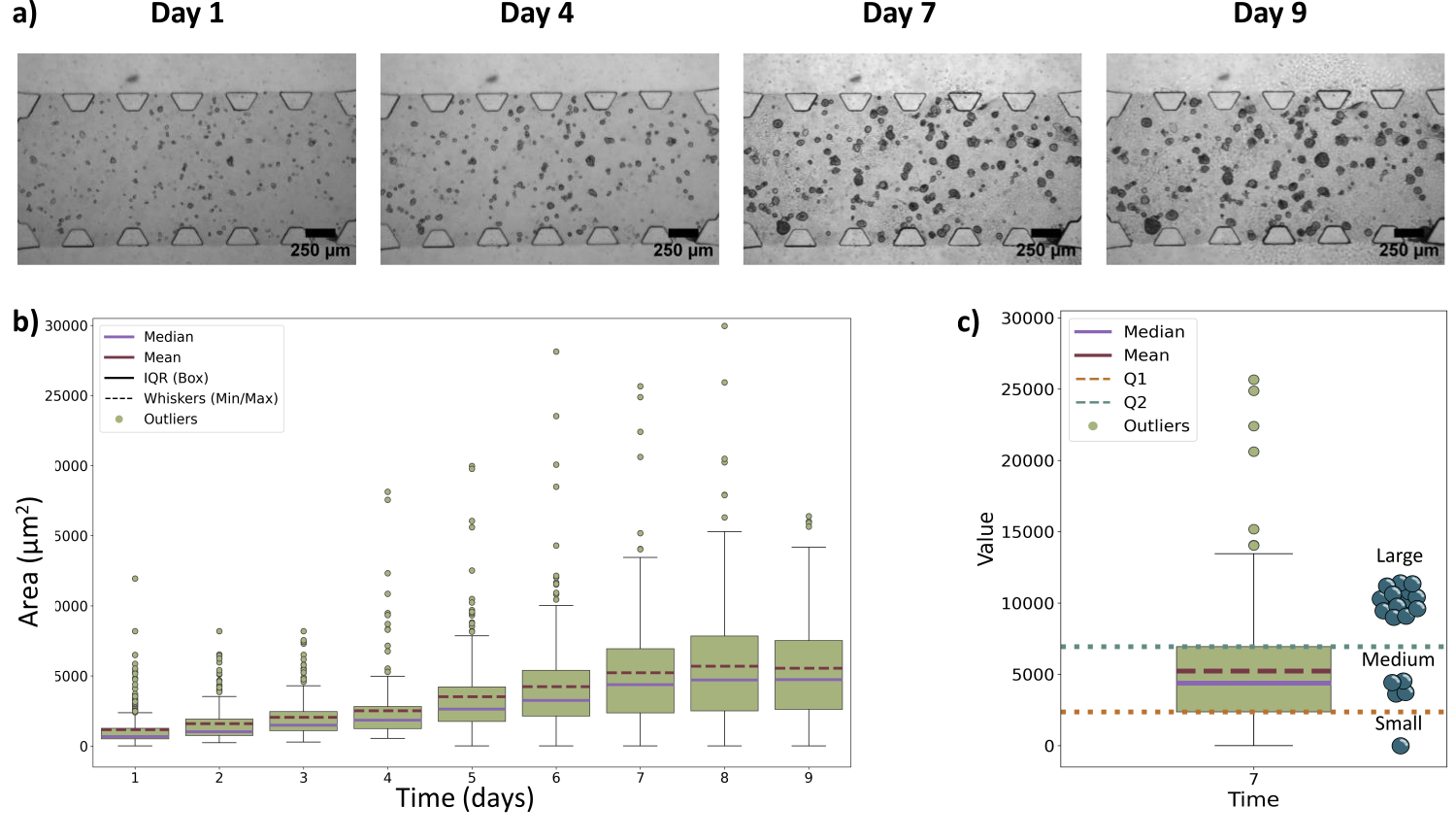}
    	\caption{Experimental results for the 3D tumour spheroids growth (4 replicates). a) Microscopy images 			of the microfluidic devices obtained as explained in section 2.2 in a Leica DM IL 			Led microscope for days 1, 4, 7 and 9. b) Evolution of the area of the organoids 			of the four experiments considered from day 1 to 9. c) Areas of the spheroids at 			day 7. For the calibration of the parameters of our computational model, we have 			divided the experimental data into three different groups due to the different 				dynamics observed in the experimental data. The division was based on using the 				quantiles of the area of the spheroids at day 7.}
    
    \label{fig:exp}
\end{figure}

Over the days of the experiment, cells progressively proliferated, forming compact spheroids by day 4. These spheroids continued to grow until day 9, generating a heterogeneous population with a broad distribution of sizes. Image segmentation and processing (as described in Section \ref{sec:image_acq}) provided quantitative measurements of spheroid area (Figure \ref{fig:exp}). From these data, a growth curve was obtained showing a steady increase in spheroid size from day 1 to day 8, with growth appearing to stabilize by day 9.\\

The temporal analysis also revealed that spheroid sizes were relatively homogeneous in the early days, whereas variability increased markedly from day 4 onwards, leading to a broad distribution. By day 9, the dispersion decreased again, producing a more compact size range. Consistently, the mean spheroid area was larger than the median at each time point, indicating the presence of numerous small spheroids alongside a subset of much larger ones, which acted as outliers. This indicates that spheroids within the same chip can display very different growth behaviours. Due to this, in the subsequent computational calibration, we divided our experimental data into three different groups based on the quantiles of the spheroid size distribution at day 7. This time point was chosen because it already showed clear differences in spheroid behaviour. Specifically, the first ($\approx 2500~\mu m^2$) and the third ($\approx 7500~\mu m^2$) quantiles were used to establish thresholds: spheroids below the first quantile were considered small, those between the first and third quantiles medium, and those above the third quantile large (Figure \ref{fig:marg}c). This approach provided a data-enhanced means of classifying spheroids while accounting for the inherent variability in their growth.

\subsection{Uncertainty quantification and parameter influence}
\begin{figure}[h]
    \centering
    \includegraphics[width=1\textwidth]{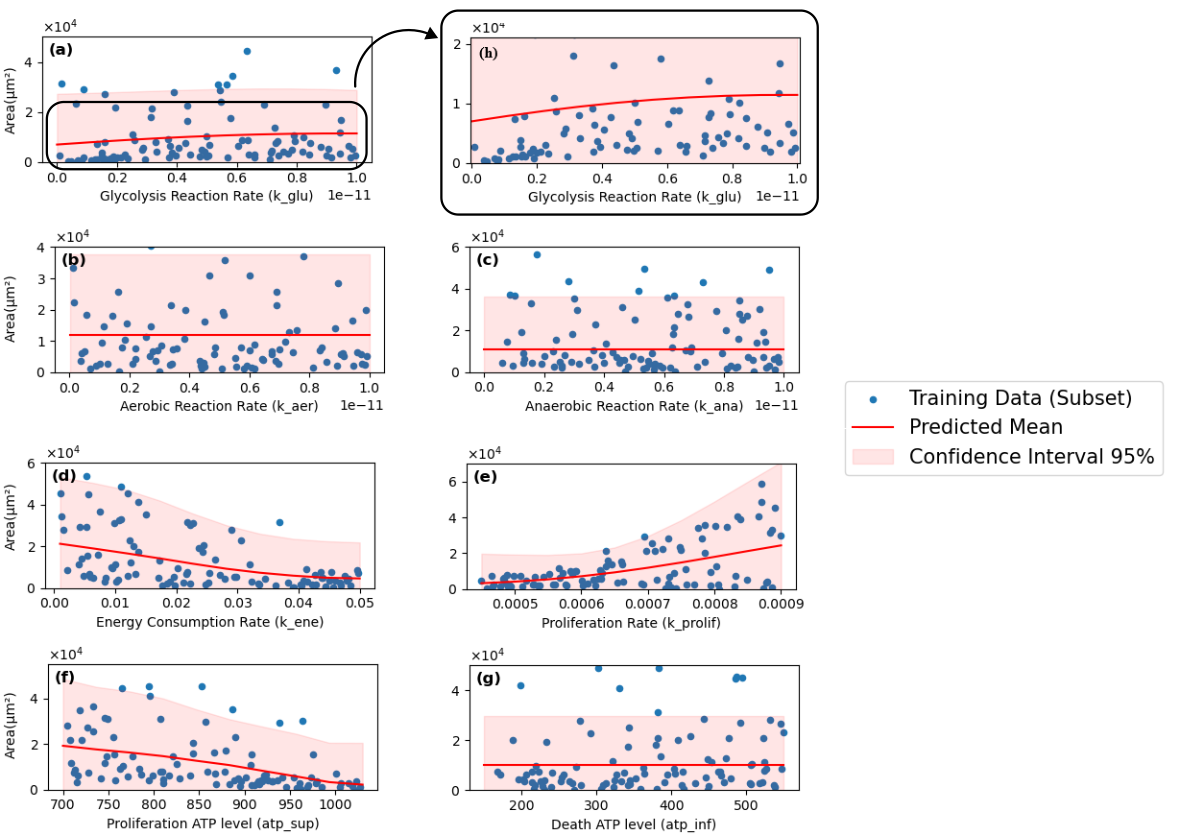}
    \caption{Scatter plots showing the impact the metabolic and non-metabolic parameters have on the target variable (Area). The red line represents the Gaussian Process predictive mean, while the shaded pink region corresponds to the 95$\%$ confidence interval obtained through marginalization over the parameter of each subplot. Blue dots represent 15\% of the training data used for inference. h) Subplot is a zoom in to see more clearly the dependency of the output of the model with $k_{glu}$ represented in the a) subplot.}
    \label{fig:marg}
\end{figure}

A scatter plot of the marginalization process was perform to visually check the impact of the different paramets on the quantity of interest. Figure \ref{fig:marg} illustrates the marginalization results for each parameter of the model (described in Section \ref{sec:stat_methods}). The blue dots of the graph are part of the training data of the GP ($D_{training}$). The red line represents the result of the marginalization process explained in equation \ref{eq:marg}.\\

Figure \ref{fig:marg} indicates that three parameters significantly influence spheroid area: energy consumption rate, proliferation rate, and proliferation ATP level. The glycolysis reaction rate exhibits a weaker, though still perceptible, influence on spheroid area, more clearly visualized in Figure \ref{fig:marg}.b). It can be seen that among the non-metabolic parameters, the proliferation rate and the proliferation ATP threshold have the most substantial influence on spheroid area. This suggests that once spheroid cells transition to a quiescent state, they are unable to revert to proliferation, ultimately leading to cell death. This hypothesis is supported by the negligible effect of the death ATP level on spheroid area. Among metabolic parameters, the energy consumption rate is the dominant factor. A minor effect of the glycolysis reaction rate is also observed (Figure \ref{fig:marg}.h). The lack of dependency on aerobic and anaerobic reaction rates suggests that any of these pathways are preferentially activated under the experimental conditions used for training the GP model.\\

\begin{figure}[h]
    \centering
    \includegraphics[width=0.5\textwidth]{"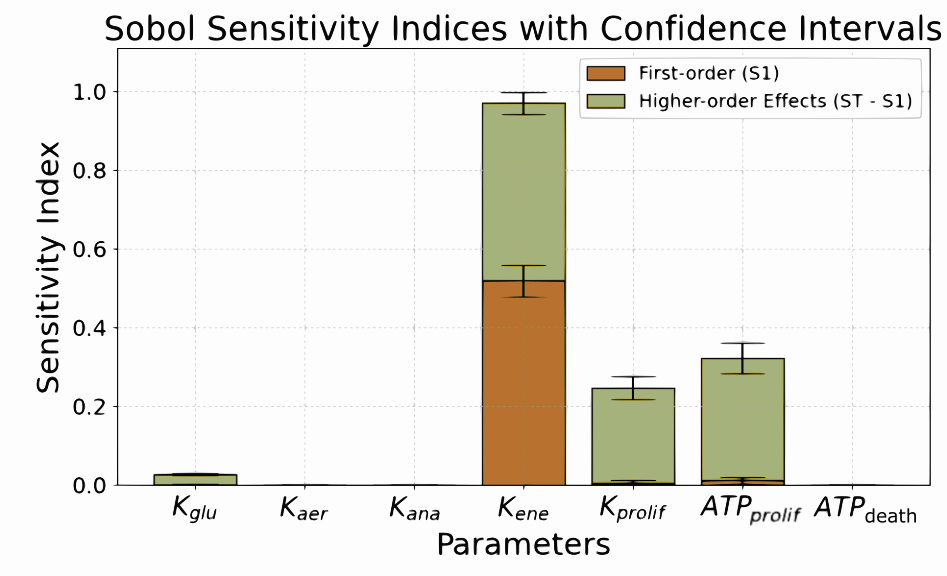"}
    \caption{ First and total order sobol indexes of the different parameters of the model at day 7. The first order sobol indexes (orange) indicates the number 		direct influence of the parameter in the output of the model. The total order sobol index (ligth green) index indicates the action of all the crossed relations between the parameters in the output of the model. }
    \label{fig:sobol_indexes}
\end{figure}

To extend the sensitivity analysis, we computed the Sobol indices of the model parameters, as indicated in Section \ref{subsec: parames_importance}. In  Figure~\ref{fig:sobol_indexes}, first and total order Sobol indexes are presented. This figure supports the conclusions drawn from Figure~\ref{fig:marg}, while also offering a more detailed understanding of the relative influence of each parameter and their interactions. Figure \ref{fig:sobol_indexes} clearly show that the energy consumption rate ($K_{ene}$) is the most influential parameter, with a substantial first-order Sobol index. When considering total-order indexes, additional dependencies emerge. Second-order Sobol indexes, which quantify pairwise parameter interactions can be found in Appendix \ref{appendix:sobol}. These results indicate that $k_{glu}$, $k_{prolif}$, and $ATP_{prolif}$ contribute meaningfully to the model output when combined with $K_{ene}$. Conversely, the other three parameters ($K_{aer}$, $K_{ana}$, and $ATP_{death}$) exhibit negligible influence and will henceforth be fixed at the midpoint of their respective range of values (Table \ref{table:final_params}).\\

To enable efficient simulation for Bayesian parameter calibration, the efficienty of the GP has been analyised. The model was trained on 1024 measurements, and achieved a Nash-Sutcliffe efficiency coefficient (Eq. \ref{eq:nash}) of $Q^2 = 0.98$, confirming its high predictive accuracy.
\subsection{Bayesian calibration of model parameters and validation against experimental data}
As described in subsection \ref{subsec:bayesian}, the Bayesian optimization process provides a probability distribution for the model parameters. The Bayesian optimization algorithm explores the parameter space defined by the prior distributions and, after comparing the GP predictions with the experimental data, yields posterior distributions of the parameters of the model.\\

\begin{figure}[h]
    \centering
    \includegraphics[width=0.9\textwidth]{"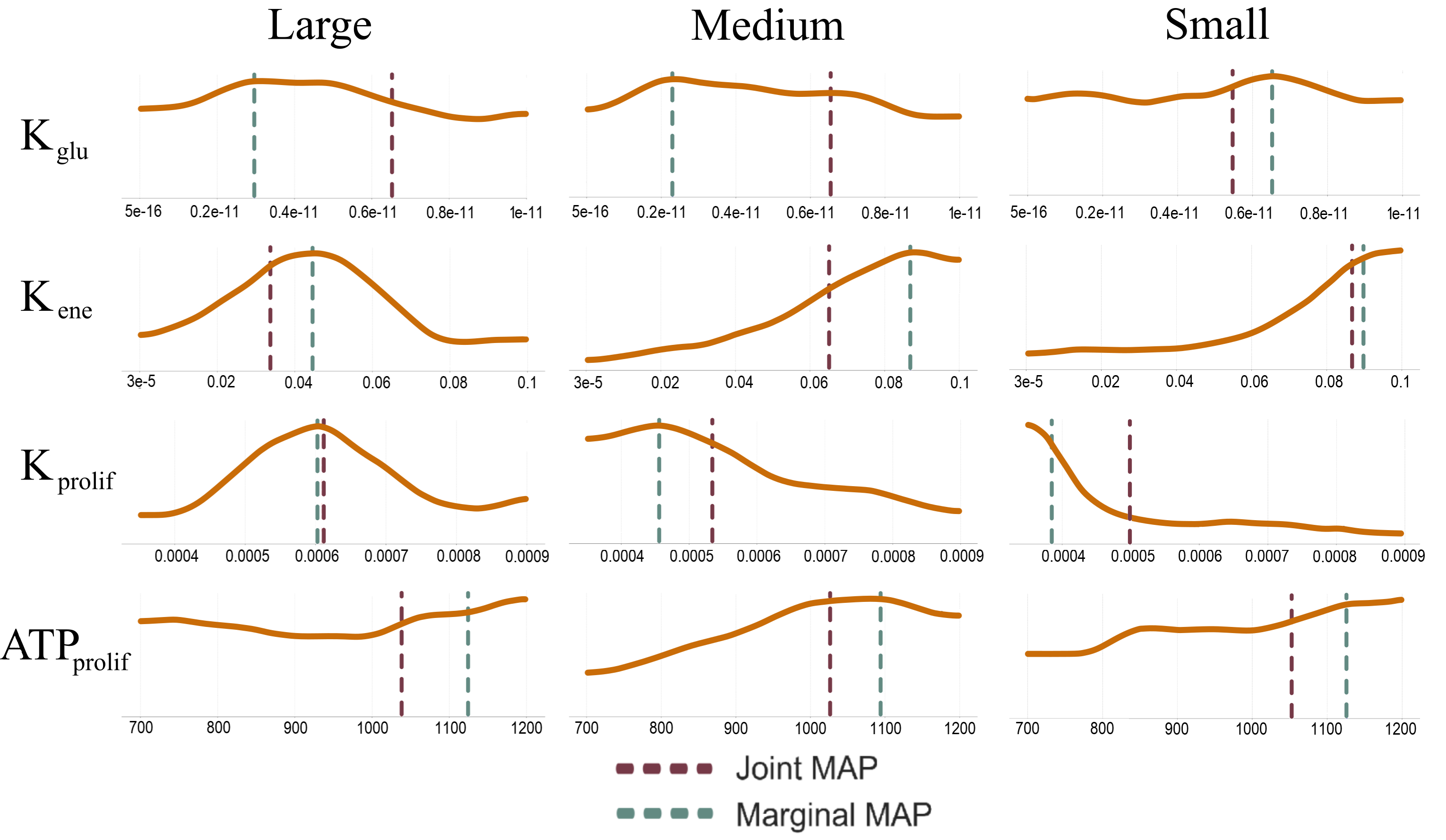"}
    \caption{Posterior distributions of the parameters for the three experimental spheroid groups (large, medium and small). The red line indicates the joint MAP for all four parameters in each group, while the green line indicates the marginal MAP for each parameter.}
    \label{fig:posteriors}
\end{figure}

\begin{figure}[h]
    \centering
    \includegraphics[width=0.65\textwidth]{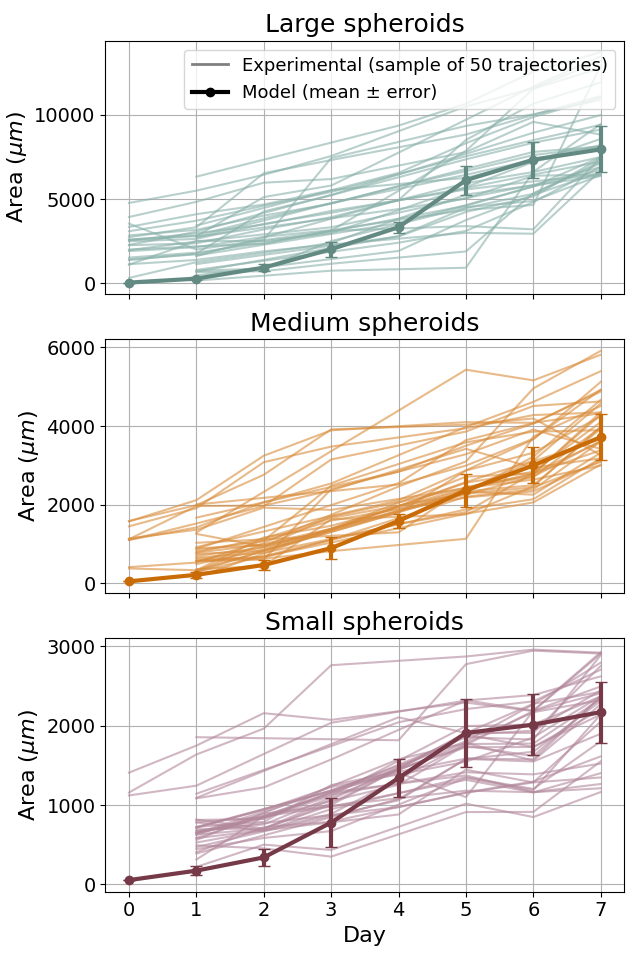}
    \caption{Computational predictions (three replicates) using the joint MAP values from Figure \ref{fig:posteriors}, compared with a subset of 35 experimental samples from day 0 to day 7. The simulations capture the main trends observed in the experimental data.}
    \label{fig:predicciones}
\end{figure}

Because the experiments in subsection \ref{subsec:experiments} revealed markedly different dynamics and regimes, the spheroids were classified into three groups, each displaying distinct behaviours. As outlined in Section \ref{sec:experiments}, parameter estimation was performed independently for each spheroid group. Uniform priors were assigned to the four parameters identified as most influential by the sensitivity analysis (Figure \ref{fig:sobol_indexes}). The prior ranges, reported in Table \ref{table:priors}, were determined either from preliminary model simulations or from values reported in the literature. In addition to the full posterior distributions, the maximum a posteriori (MAP) estimate was computed for each experimental group. This estimate corresponds to the parameter set that best reproduces the experimental measurements. Figure \ref{fig:posteriors} shows the posterior distributions of the four parameters, together with the marginal MAP values (estimated individually for each parameter) and the joint MAP values (estimated simultaneously for all parameters). The figure highlights clear differences both among the posterior distributions of the four parameters and across the three spheroid groups.    In particular, the posterior distributions of $K_{ene}$ and $K_{prolif}$ are more sharply defined than those of $K_{glu}$ and $ATP_{prolif}$. Furthermore, differences between groups can be observed both in the posterior distributions and in the MAP estimates. Specifically, larger spheroids exhibit higher $K_{prolif}$ values and lower $K_{ene}$ values compared with smaller spheroids. Once the joint MAP values were obtained, they were introduced into the computational model to evaluate whether the simulations could reproduce the experimental observations (Table \ref{table:final_params}). Three simulations were performed for each group (large, medium, and small spheroids). The resulting spheroid areas were measured, averaged across replicates, and compared with the experimental data (Figure \ref{fig:predicciones}).\\

\begin{table}[]
\centering
\begin{tabular}{cc}
\multicolumn{2}{c}{Prior distributions}                 \\ \hline
\multicolumn{1}{c|}{$k_{glu}^{*}$}       & $\mathit{U}\left[5\cdot10^{-16} - 1\cdot10^{-11}\right]$ \\ \hline
\multicolumn{1}{c|}{$k_{ene}^{*}$}       & $\mathit{U}\left[3\cdot10^{-5} - 0.1\right]$              \\ \hline
\multicolumn{1}{c|}{$k_{prolif}^{\dagger}$}    & $\mathit{U}\left[3.5\cdot10^{-4} - 9\cdot10^{-4}\right]$  \\ \hline
\multicolumn{1}{c|}{$ATP_{prolif}^{*}$}  & $\mathit{U}\left[700 - 1200\right]$  
\end{tabular}
\caption{Prior distributions used in the Bayesian optimization process. $^{*}$ Values obtained from preliminary simulations of the model. $^{\dagger}$ Values from \cite{nisar_hypoxia_2023}.}
\label{table:priors}
\end{table}

\begin{table}[h!]
\centering
\begin{tabular}{c|c|c|c}
\textbf{Parameters} &
\makecell{\textbf{Large spheroids}} &
\makecell{\textbf{Medium spheroids}} &
\makecell{\textbf{Small spheroids}} \\ \hline
$k_{\text{glu}}^{*}$    & $0.66\times10^{-11}$ & $0.66\times10^{-11}$ & $0.54\times10^{-11}$ \\
$k_{\text{aer}}^{\dagger}$    & $5\times10^{-12}$    & $5\times10^{-12}$    & $5\times10^{-12}$ \\
$k_{\text{ana}}^{\dagger}$    & $5\times10^{-12}$    & $5\times10^{-12}$    & $5\times10^{-12}$ \\
$k_{\text{ene}}^{*}$    & 0.033 & 0.065 & 0.084 \\
$k_{\text{prolif}}^{*}$ & 0.00061 & 0.00057 & 0.0005 \\
$ATP_{\text{prolif}}^{*}$ & 1040 & 1025 & 1050 \\
$ATP_{\text{death}}^{\dagger}$ & 500 & 500 & 500 \\
\end{tabular}
\caption{Parameter values used for the computational predictions of the model.$^{*}$ Obtained from the Bayesian calibration process. $^\dagger$ Value taken from the initial range of parameters. These parameters do not influence the model in this parameter range. }
\label{table:final_params}
\end{table}

Overall, the computational model successfully reproduced the experimental dynamics across the three spheroid groups. However, during the initial days, the simulated spheroid areas tended to be slightly smaller than those measured in vitro. This effect might be due to the assumption that all spheroids start being an individual cell that proliferate through the days, updating its size. However experimental measurements clearly indicate that spheroids start their growth with a size different than that of individual cells. This might be due to an error in the experiments, probably because the diffusor used did not favor single-cell deposition in the microfluidic devices correctly and small aggregates of cells formed in the microfluidic devices at the beginning of the experiments. Despite this, Figure \ref{fig:predicciones}, shows that our computational model is able to overcome this initial difference between the experimental reality and the computational approach and reproduce the general time-dependent behaviour of the experimental data, specially for the last days of the experiments. \\

\subsection{Collagen concentration regulates the size of the tumor spheroid.}
Once the parameters of our model were calibrated and the simulations successfully reproduced the behaviours observed in the experimental setup (Figure \ref{fig:predicciones}), we analysed the model response across different collagen densities.\\

\begin{figure}[h]
    \centering
    \includegraphics[width=0.7\textwidth]{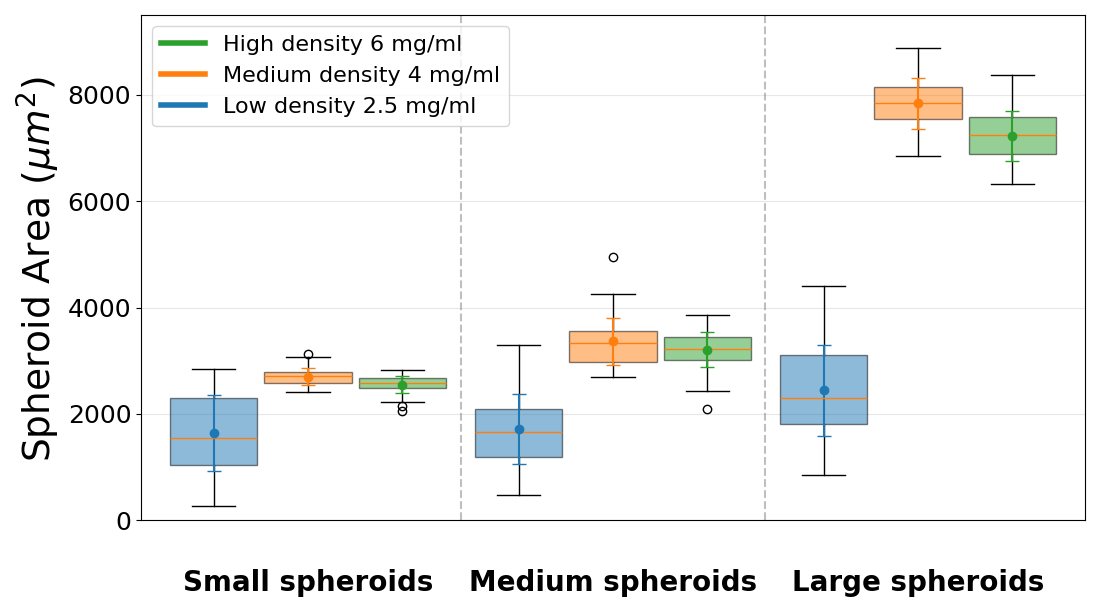}
    \caption{Spheroid areas obtained from computational simulations across three ECM collagen concentrations (2.5, 4, and 6 mg/ml). Ten replicates were performed per condition. Despite identical proliferation rates across matrices (see Appendix 1, Figure \ref{fig:ap_num_cels}), clear differences in spheroid area emerge. Low-density collagen (2.5 mg/ml) produces smaller spheroids due to fragmentation of the initial aggregate. Medium-density collagen (4 mg/ml) yields the largest spheroids, whereas high-density collagen (6 mg/ml) results in more compact, smaller aggregates. }
    \label{fig:comp_area}
\end{figure}

Using the parameters listed in Table \ref{table:final_params}, we ran 10 simulations for the three sets of parameters and for three different ECM collagen concentrations: a high-density matrix (6 mg/ml), a medium-density matrix (4 mg/ml), and a low-density matrix (2.5 mg/ml). Several measurements were obtained from the spheroids formed in the simulations. Figure \ref{fig:comp_area} shows the resulting spheroid areas. As shown, low-density collagen produces the smallest spheroids. This reduction is not due to changes in proliferation, as cell proliferation does not depend on the collagen density of the ECM (see Appendix 1). Instead, the smaller spheroids observed in the 2.5 mg/ml condition result from fragmentation of the initial spheroid into multiple smaller aggregates. This hypothesis is supported by the greater number of spheroids formed in low-density matrices (Appendix 1).

\begin{figure}[h]
    \centering
    \includegraphics[width=0.7\textwidth]{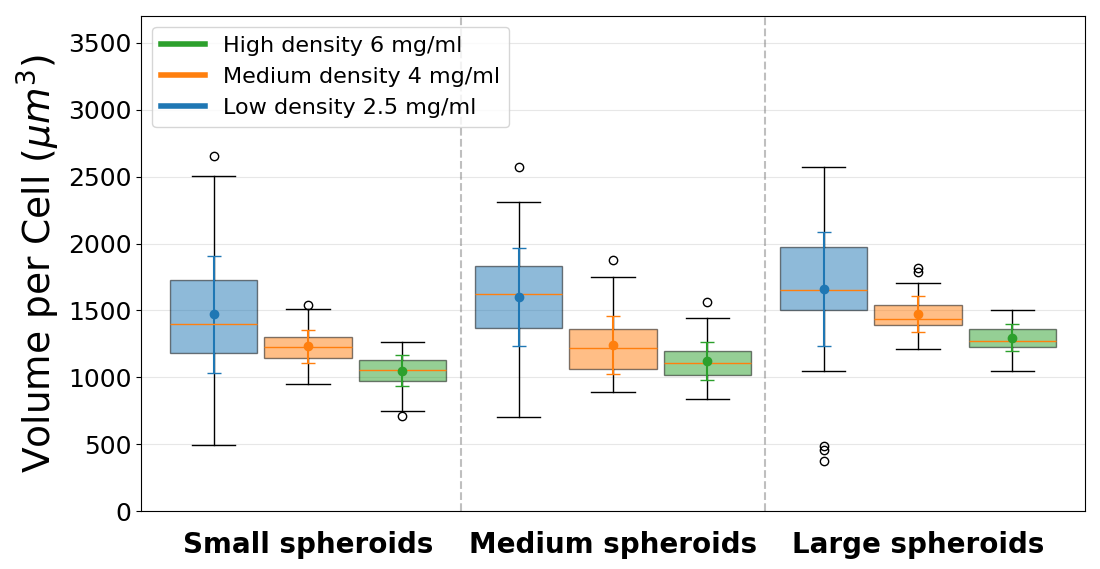}
    \caption{Mean volume occupied per cell within spheroids across ECM collagen densities. This metric reflects spheroid compactness. Cells in high-density collagen (6 mg/ml) occupy less volume on average, indicating increased compaction relative to medium- and low-density matrices.  }
    \label{fig:comp_vol_per_cell}
\end{figure}

Figure \ref{fig:comp_area} also shows that medium-density ECMs tend to yield larger spheroids than high-density ECMs. Again, this difference cannot be attributed to variations in proliferation, but rather to changes in spheroid compaction. Spheroids formed in high-density matrices are more compressed and therefore occupy a smaller area. To verify this behaviour, we quantified the mean volume occupied per cell within each spheroid. The results (Figure \ref{fig:comp_vol_per_cell}) support this interpretation: cells in high-density matrices have less available space, resulting in more compact spheroids. These findings confirm that differences in spheroid compactness drive the size differences between medium- and high-density matrices, where the number of spheroids formed is similar (Appendix 1).\\

\subsection{Epithelial-like behaviour prevails over the role of the ECM}\label{sec:mesen_epi}
According to Hernández-Hatibi et al. \cite{hernandez-hatibi_quantitative_2025}, mesenchymal cell lines such as Panc-1 and MIA PaCa-2 tend to exhibit behaviours consistent with our simulations: spheroids grown in 4 mg/ml collagen form larger structures than those grown in 6 mg/ml matrices. In contrast, more epithelial cell lines (BxPC-3 and Capan-2) do not show this behaviour. These lines show either no significant differences in spheroid size across ECM densities or exhibit larger spheroids in denser matrices rather than in the 4 mg/ml conditions. This behaviour could be explained by the greater motility and metastatic potential of mesenchymal cell lines. In motility-driven spheroid expansion, as it is in this case where all the conditions present the same proliferation (See Appendix 1), increased ECM density restricts cell movement, thus limiting spheroid growth. Conversely, epithelial cells exhibit significantly lower motility; therefore, their spheroid expansion is governed primarily by proliferation rather than movement. As a result, epithelial spheroids are less sensitive to increased ECM stiffness and the associated reduction in cell motility. To test this hipothesis in our model, we simulated epithelial-like behaviour by reducing the locomotive force (Equation \ref{eq:force}) of computational cells by a factor of 0.1, and we quantified spheroid area under this reduced-velocity condition. We then performed an statistical analysis (See  Appendix \ref{appendix:normal_homo_anova}) on the measurements derived from the computational spheroids to asses wheter they align with the findings of Hernandez-Hatibi et al.\\

\begin{figure}[h]
    \centering
    \includegraphics[width=1\textwidth]{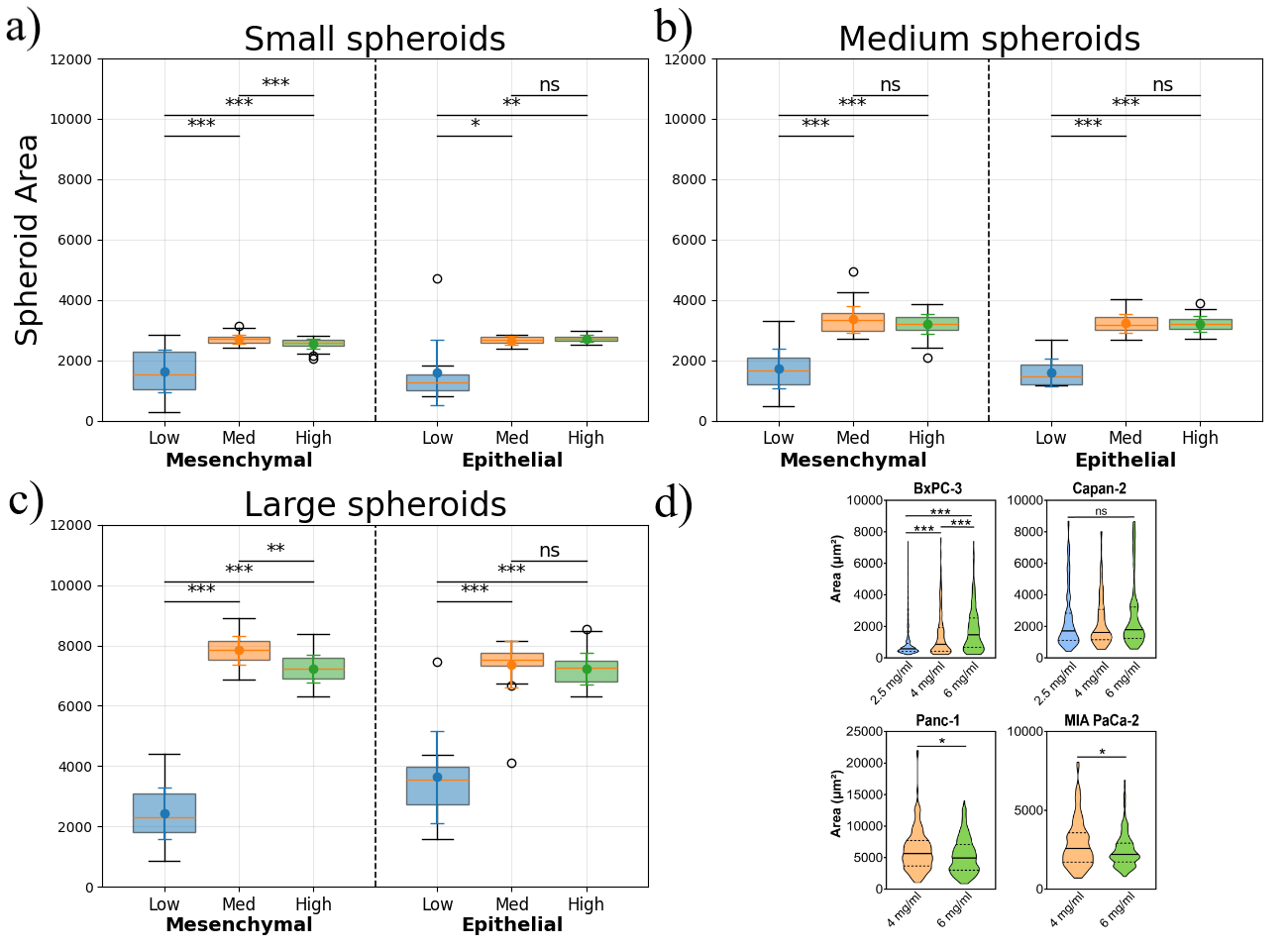}
    \caption{a), b) and c) Comparation between the spheroid areas from mesenchymal and epithelial cell lines. A statistical analysis was conducted to evaluate the significance of the differences in size observed between the groups. The detailed methodology is provided in Appendix \ref{appendix:normal_homo_anova}. d) Experimental results from Hernandez-Hatibi et al.\cite{hernandez-hatibi_quantitative_2025} for different mesenquimal (Panc-1 and MIA PaCa-2) and epithelial (BxPC-3 and Capan-2) cell lines. Under reduced motility, spheroid sizes in medium- and high-density matrices converge, as it can be seen in the statistical analysis. This reproduces qualitativelythe behaviour reported experimentally for epithelial cell lines(d).}
    \label{fig:comp_cel_lines}
\end{figure}

 Figure \ref{fig:comp_cel_lines} compares the results and statistical analysis obtained for the spheroid area of the simulations in ephithelial and mesenchymal cell lines. As observed in Figure \ref{fig:comp_cel_lines}, reducing cell velocity results in the disappearance of size differences between spheroids grown in high- and medium-density matrices. This difference can be appreciated not only visually but also through the statistical analysis presented in the figure. Therefore, these simulations  reproduce the behaviour reported by Hernández-Hatibi et al. \cite{hernandez-hatibi_quantitative_2025} for epithelial cell lines, confirming our hypothesis, reduced movility reproduce epithelial behaviours. \\

\section{Discussion } \label{sec:conclusions}

Computational models are powerful tools to complement and improve the understanding of in vivo and in vitro experiments. Nevertheless, current models still face mayor challenges in order to become fully realistic and broadly applicable. In particular, they must  capture biological behaviour in its entirety, rather than focusing on isolated aspects, and reproduce experimental results not only qualitatively but also quantitatively. In this study, we take a step toward addressing both of these challenges. \\

We present a computational model that integrates metabolic and mechanical effects, thereby bridging two approaches that are often studied separately. This model enables the evaluation of diverse biological quantities related to metabolism and mechanics, such as metabolites consumtion and secretion or the  velocities of the cells. Simultaneously, it provides multicellular structural descriptors, including the number and size of spheroids formed, thereby offering insight into the organization and development of tumour-like structures. Building on this modelling framework, the sensitivity analysis and Bayesian parameter calibration performed in this work enables to infer not only qualitative but also quantitative results from our model. From Figures \ref{fig:marg} and \ref{fig:sobol_indexes}, we infer that within the explored parameter space and experimental conditions, three parameters ($k_{aer}$, $k_{ana}$, and $ATP_{death}$) present negligible influence on the model output.  This suggests that under the specific conditions tested, for the A549 cells both the aerobic and anaerobic metabolic pathways are balanced and cells do not show a clear preference for either pathway. Nevertheless, a broader sensitivity analysis would be required to confirm that these parameters remain negligible across wider parameter ranges and experimental contexts. The posterior distributions obtained through Bayesian calibration (Figure \ref{fig:posteriors}) indicate that larger spheroids are characterised by higher proliferation rates ($k_{prolif}$) and lower energy consumption ($k_{ene}$). This suggests that spheroids tend to grow larger when they achieve greater energy efficiency, meaning they can proliferate more rapidly while expending less energy. As shown in Figure \ref{fig:predicciones}, combining our computational model with the Bayesian calibration pipeline enabled the quantitative reproduction of experimental results obtained in vitro using microfluidic devices. This agreement between simulations and experiments reinforces the model’s capacity to capture key biophysical mechanisms underlying spheroid development.\\

Moreover, our model is not only able to reproduce experimental data, but also to generate mechanistic insights into spheroid growth under different microenvironmental conditions. As illustrated in Figure \ref{fig:comp_area}, simulations reveal that spheroids reach their largest sizes at intermediate collagen densities (4 mg/mL). This behaviour arises because low-density matrices fail to physically confine cells, leading to the formation of multiple smaller spheroids, whereas high-density matrices impose strong mechanical constraints that compact the structure and reduce the available space per cell. These findings highlight the model’s capacity to connect extracellular matrix properties with emergent multicellular organisation. Additionally, the model proved to be flexible enough to capture qualitative differences between cell lines and phenotypes. By modifying the parameters that determine cell motility, we were able to reproduce behaviours typically associated with epithelial and mesenchymal states (Section \ref{sec:mesen_epi}). This demonstrates that the computational framework can be tuned to represent diverse biological scenarios and suggests its potential for studying transitions between phenotypic states or comparing different tumour cell populations.\\

Despite the model's success in capturing experimental data (Figure \ref{fig:predicciones}), certain limitations remainand open important avenues for future research. First, our metabolic model focus on the glycolysis-OXPHOS axis. However, recent studies have demonstrated the importance of the role of other metabolites for the production of energy. Guerrero-Lopez et al  demonstrated that glutamine metabolism is crucial in 3D cultures, especially under glucose restriction \cite{guerrero_lopez_2d_2025}. The future incorporation of glutamine pathways will increase the model's biological realism. The ECM is currently simulated as a homogeneous continuum medium that provides mechanical resistance. However, the ECM consists of heterogeneous fibre networks that directly influence tumour cell migration, and tumour cells can, in turn, remodel these fibres during growth and invasion \cite{noel_physimess_2024, rejniak_cell-ecm_2016}.  While our continuum approximation is computationally feasible and sufficient to reproduce certain experimental observations \cite{goncalves_extracellular_2021}, it does not capture ECM heterogeneity or its dynamic interactions with tumour cells.\\

Despite these limitations, the proposed model has proven to be a robust tool for reproducing in vitro experiments and for generating biologically meaningful insights into the processes governing cancer cell behaviour and tumour spheroid formation. In addition to that, the model here proposed fills the current gap in the literature of a single agent-based cancer model that couples the metabolic and mechanic aspects of tumour spheroid formation.  This work thus contributes to the development of integrated computational frameworks that can bridge metabolism and mechanics, supporting both experimental research and the advancement of predictive oncology.\\

\section{Acknowledgements}

This work is part of the project PID2021-122409OB-C21 and PID2024-155384OB-C21 funded by 
MICIU/AEI/10.13039/501100011033 and
by “ERDF/EU”. P.G-G. gratefully acknowledges the Spanish Government for an FPI predoctoral contract (PRE2022-104205). P.G-L.gratefully acknowledges the support of the Government of Aragon (Grant No 2021-25). S. H-R. gratefully acknowledges the support of Fundación Ibercaja-CAI (IT 6/24) and the Alexander von Humboldt Foundation and the Carl Friedrich von Siemens Foundation for their support through the Alexander von Humboldt Postdoctoral Fellowship Program. JM.G-A. gratefully acknowledges the support of ERC-2020-Advanced GA Nr. 101018587 (ICoMICS)

\section{Author contribution}

Conceptualization: P. G-G., S. H-R. and JM. G-A.; Computational design: P. G-G. and S.H-R; Computational implementation: P G-G; Experimental implementation: P. G-L.; Funding acquisition: JM. G-A.; Original draft: P. G-G.; Review and editing: P. G-G., P. G-L., S. H-R. and JM. G-A.

\printbibliography
\appendix
\newpage
\section{Appendix 1. First, second and total order Sobol indexes.}\label{appendix:sobol}
In this appendix, we present the results of the Sobol sensitivity analysis performed for our computational model. Table \ref{table:1sobol_index} reports the first-order and total-order Sobol indices, which quantify the individual contribution of each parameter and their combined (global) influence on model output, respectively. Table \ref{table:sij_matrix} shows the second-order Sobol indices, providing insight into pairwise parameter interactions within the model.

\begin{table}[h!]
\centering
\begin{tabular}{c|c|c}
\textbf{Parameters} & \textbf{$S_i^1$} & \textbf{$S_i^{TOT}$} \\ \hline
$k_{glu}$           & 0.0002         & 0.03                 \\
$k_{aer}$           & 0.0            & 0.0                  \\
$k_{ana}$           & 0.0            & 0.0                  \\
$k_{ene}$           & 0.5191         & 0.97                 \\
$k_{prolif}$        & 0.006          & 0.24                 \\
$ATP_{prolif}$         & 0.0113         & 0.32                 \\
$ATP_{death}$         & 0.0            & 0.0                 
\end{tabular}
\caption{First and total order Sobol indexes of the different parameters of the model}
\label{table:1sobol_index}
\end{table}


\begin{table}[h]
\centering
\renewcommand{\arraystretch}{1.4}
\setlength{\tabcolsep}{10pt}
\begin{tabular}{|l||c|c|c|c|c|c|}
\hline
\textbf{$S_{ij}^2$} & $k_{aer}$ & $k_{ana}$ & $k_{ene}$ & $k_{prolif}$ & $ATP_{prolif}$ & $ATP_{death}$ \\
\hline\hline
$k_{glu}$     & 0.0025 & 0.0025 & 0.0159 & 0.0024 & 0.0015 & 0.0025 \\ \hline
$k_{aer}$     &        & 0.0000 & 0.0000 & 0.0000 & 0.0000 & 0.0000 \\ \hline
$k_{ana}$     &        &        & 0.0000 & 0.0000 & 0.0000 & 0.0000 \\ \hline
$k_{ene}$     &        &        &        & 0.1620 & 0.2087 & 0.0061 \\ \hline
$k_{prolif}$  &        &        &        &        & 0.0031 & 0.0019 \\ \hline
$ATP_{prolif}$ &        &        &        &        &        & 0.0035 \\ \hline
\end{tabular}
\caption{Squared sensitivity matrix $S_{ij}^2$ for parameter pairs $(i,j)$. Only upper-triangular values shown.}
\label{table:sij_matrix}
\end{table}
\section{Appendix 2. Additional measurements and statitical analysis for different collagen density.  } \label{appendix:add_measurements}
In this appendix we present some extra measurements performed for the parameters obtained through the Bayesian optimization process.

\begin{figure}[h]
    \centering
    \includegraphics[width=0.7\textwidth]{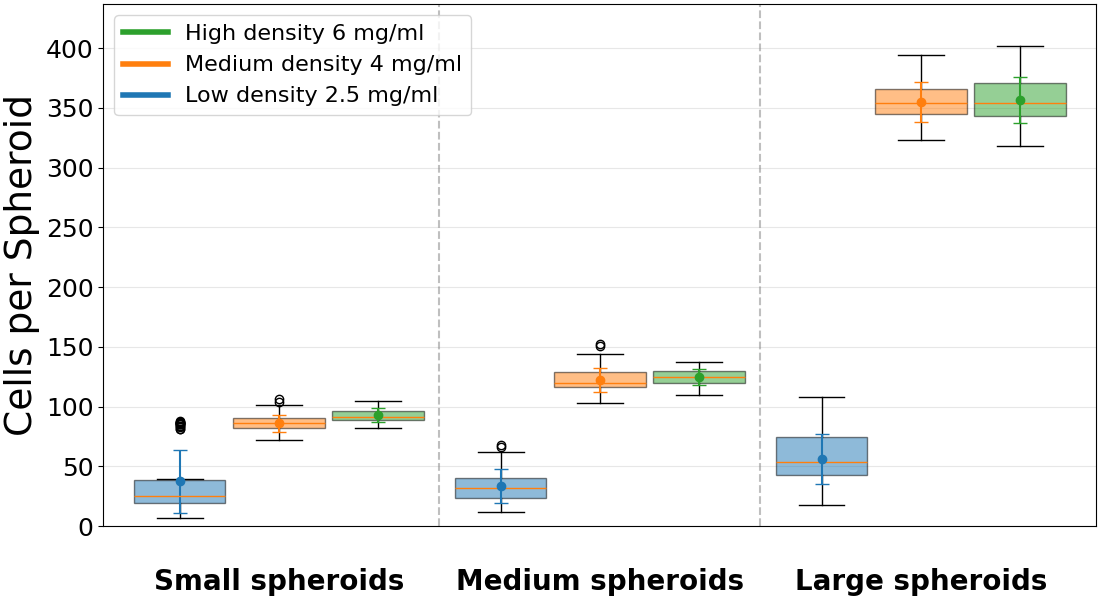}
    \caption{ Cells per spheroid. Spheroids formed in low-density collagen ECM contain significantly fewer cells, as they tend to fragment under these conditions.} 
    \label{fig:ap_cells_per_sphe}
\end{figure}

\begin{figure}[h]
    \centering
    \includegraphics[width=0.7\textwidth]{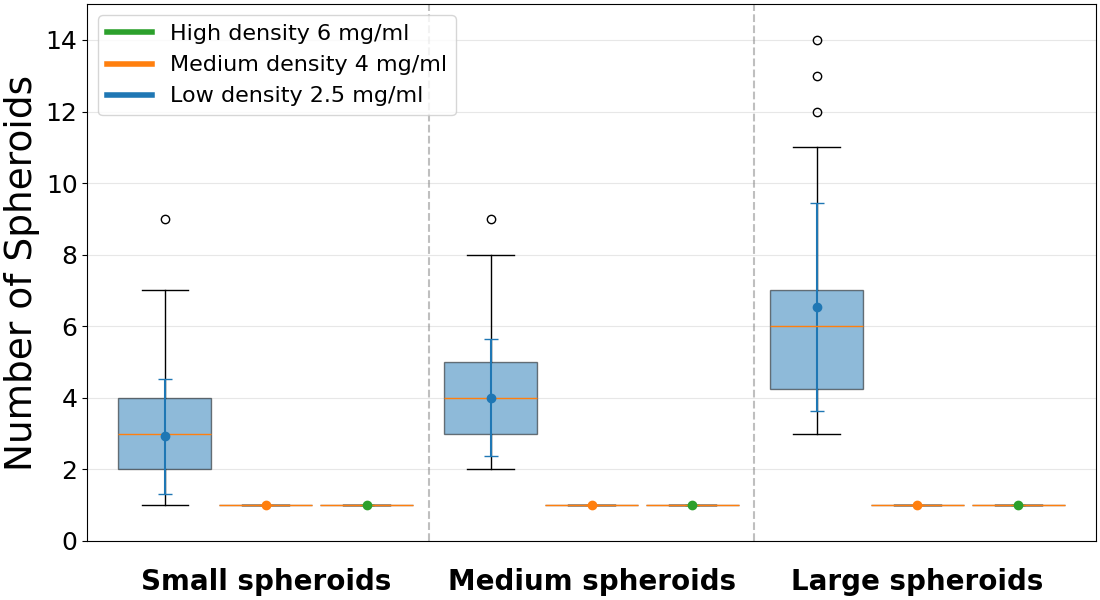}
    \caption {Number of spheroids. Spheroids formed in low-density collagen ECM tend to fragment, and therefore, the number of total spheroids is bigger for low collagen density.}
    \label{fig:ap_num_spheroids}
\end{figure}

\begin{figure}[h]
    \centering
    \includegraphics[width=0.7\textwidth]{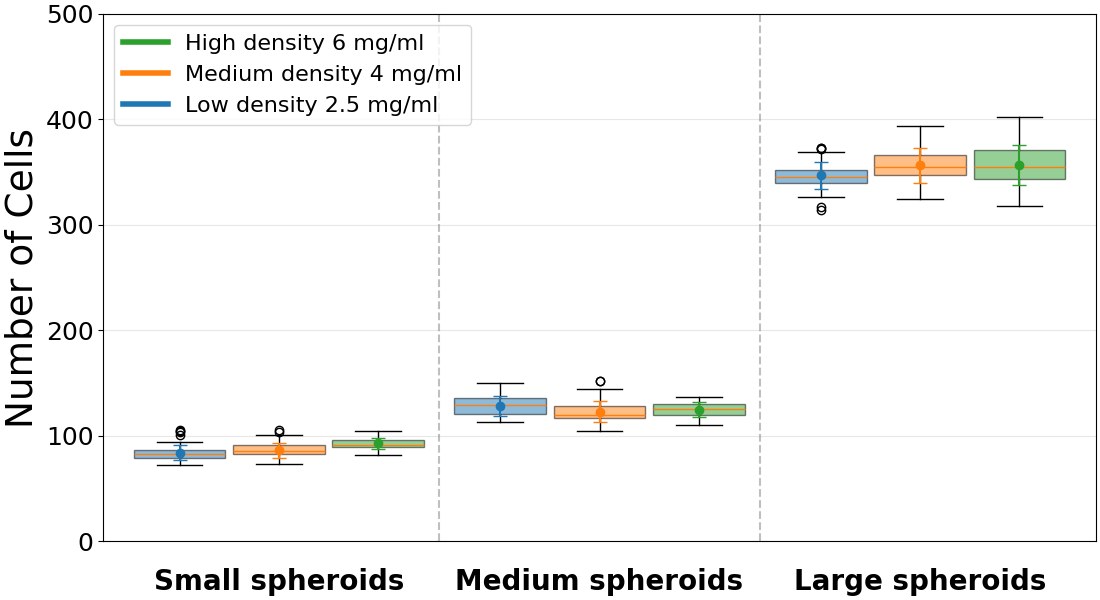}
    \caption{Number of cells. Proliferation is not affected by the ECM collagen density.}
    \label{fig:ap_num_cels}
\end{figure}
\newpage
\newpage

\section{Appendix 3.  Normality analysis and homogeneity of the variances check}\label{appendix:normal_homo_anova}

To evaluate whether the differences observed across simulation scenarios were statistically significant, we applied either an ANOVA or a Kruskal–Wallis test, selecting the appropriate method based on the normality and homogeneity of variance of the data. \\

Normality was assessed using the Shapiro–Wilk test \cite{shapiro_analysis_1965}, which is well suited for small sample sizes and therefore appropriate for our dataset, composed of fewer than 50 simulations per condition. The outcomes of the normality analyses are provided in Tables \ref{table:normality_check} and \ref{table:normality_check_vel_red}.\\

Homogeneity of variances across groups was subsequently evaluated using Levene’s test, with results reported in Tables \ref{table:levene} and \ref{table:levene_vel_red}.\\

Based on the combined results of the normality and variance-homogeneity assessments, each dataset was assigned to either a parametric ANOVA followed by a Tukey–Kramer post-hoc or a non-parametric Kruskal–Wallis test with post hoc Dunn’s test for further statistical comparison. The corresponding decisions for each case are summarized in Tables \ref{table:kruskal_anova} and \ref{table:kruskal_anova_red}.\\

\begin{table}[h!]
\centering
\begin{tabular}{cc|c|c|c|c|c|}
\cline{3-7}
                                                                                                   &                                                          & Area                                                  & \begin{tabular}[c]{@{}c@{}}Number of \\ spheroids\end{tabular} & \begin{tabular}[c]{@{}c@{}}Cells per\\ spheroid\end{tabular} & \begin{tabular}[c]{@{}c@{}}Number\\ of cells\end{tabular} & \begin{tabular}[c]{@{}c@{}}Volume\\ per cell\end{tabular} \\ \hline
\multicolumn{1}{|c|}{}                                                                             & \begin{tabular}[c]{@{}c@{}}High\\ density\end{tabular}   & \cellcolor[HTML]{9AFF99}0.3210                        & \cellcolor[HTML]{9AFF99}1                                     & \cellcolor[HTML]{9AFF99}0.2350                               & \cellcolor[HTML]{9AFF99}0.2350                            & \cellcolor[HTML]{9AFF99}0.8732                            \\ \cline{2-7} 
\multicolumn{1}{|c|}{}                                                                             & \begin{tabular}[c]{@{}c@{}}Medium\\ density\end{tabular} & \cellcolor[HTML]{9AFF99}0.3716                        & \cellcolor[HTML]{9AFF99}1                                     & \cellcolor[HTML]{9AFF99}0.3864                               & \cellcolor[HTML]{9AFF99}0.1980                            & \cellcolor[HTML]{9AFF99}0.4070                            \\ \cline{2-7} 
\multicolumn{1}{|c|}{\multirow{-3}{*}{\begin{tabular}[c]{@{}c@{}}Small\\ Spheroids\end{tabular}}}  & \begin{tabular}[c]{@{}c@{}}Low\\ density\end{tabular}    & \cellcolor[HTML]{9AFF99}0.3594                        & \cellcolor[HTML]{FFCCC9}0.0005                                & \cellcolor[HTML]{9AFF99}0.8110                               & \cellcolor[HTML]{9AFF99}0.0522                            & \cellcolor[HTML]{9AFF99}0.4769                            \\ \hline
\multicolumn{1}{|c|}{}                                                                             & \begin{tabular}[c]{@{}c@{}}High\\ density\end{tabular}   & \cellcolor[HTML]{9AFF99}0.7727                        & \cellcolor[HTML]{9AFF99}{\color[HTML]{333333} 1}              & \cellcolor[HTML]{9AFF99}0.8421                               & \cellcolor[HTML]{9AFF99}0.8421                            & \cellcolor[HTML]{9AFF99}0.2879                            \\ \cline{2-7} 
\multicolumn{1}{|c|}{}                                                                             & \begin{tabular}[c]{@{}c@{}}Medium\\ density\end{tabular} & \cellcolor[HTML]{9AFF99}0.3354                        & \cellcolor[HTML]{9AFF99}{\color[HTML]{333333} 1}              & \cellcolor[HTML]{9AFF99}0.8314                               & \cellcolor[HTML]{9AFF99}0.8327                            & \cellcolor[HTML]{9AFF99}0.2142                            \\ \cline{2-7} 
\multicolumn{1}{|c|}{\multirow{-3}{*}{\begin{tabular}[c]{@{}c@{}}Medium\\ Spheroids\end{tabular}}} & \begin{tabular}[c]{@{}c@{}}Low\\ density\end{tabular}    & \cellcolor[HTML]{9AFF99}0.4032                        & \cellcolor[HTML]{9AFF99}0.5530                                & \cellcolor[HTML]{9AFF99}0.0861                               & \cellcolor[HTML]{FFCCC9}0.0099                            & \cellcolor[HTML]{9AFF99}0.8454                            \\ \hline
\multicolumn{1}{|c|}{}                                                                             & \begin{tabular}[c]{@{}c@{}}High\\ density\end{tabular}   & \cellcolor[HTML]{9AFF99}0.65791                       & \cellcolor[HTML]{9AFF99}1                                     & \cellcolor[HTML]{9AFF99}0.7715                               & \cellcolor[HTML]{9AFF99}0.7715                            & \cellcolor[HTML]{9AFF99}0.1724                            \\ \cline{2-7} 
\multicolumn{1}{|c|}{}                                                                             & \begin{tabular}[c]{@{}c@{}}Medium\\ density\end{tabular} & \cellcolor[HTML]{9AFF99}0.6332                        & \cellcolor[HTML]{9AFF99}1                                     & \cellcolor[HTML]{9AFF99}0.4411                               & \cellcolor[HTML]{9AFF99}0.4597                            & \cellcolor[HTML]{9AFF99}0.9095                            \\ \cline{2-7} 
\multicolumn{1}{|c|}{\multirow{-3}{*}{\begin{tabular}[c]{@{}c@{}}Big\\ Spheroids\end{tabular}}}    & \begin{tabular}[c]{@{}c@{}}Low\\ density\end{tabular}    & \cellcolor[HTML]{FFCCC9}{\color[HTML]{333333} 0.0193} & \cellcolor[HTML]{9AFF99}0.9925                                & \cellcolor[HTML]{FFCCC9}0.0415                               & \cellcolor[HTML]{9AFF99}0.2164                            & \cellcolor[HTML]{9AFF99}0.4705                            \\ \hline
\end{tabular}
\caption{This data shows the $p-values$ of the Shapiro-Wilk test of normality for the computational simulations of Figure \ref{fig:comp_area} and \ref{fig:comp_vol_per_cell}. The green colored boxes  follow normal distributions. The red colored boxes do not.  }
\label{table:normality_check}
\end{table}

\begin{table}[h!]
\centering
\begin{tabular}{cc|c|c|c|c|c|}
\cline{3-7}
                                                                                                   &                                                          & Area                                                  & \begin{tabular}[c]{@{}c@{}}Number of \\ spheroids\end{tabular} & \begin{tabular}[c]{@{}c@{}}Cells per\\ spheroid\end{tabular} & \begin{tabular}[c]{@{}c@{}}Number\\ of cells\end{tabular} & \begin{tabular}[c]{@{}c@{}}Volume\\ per cell\end{tabular} \\ \hline
\multicolumn{1}{|c|}{}                                                                             & \begin{tabular}[c]{@{}c@{}}High\\ density\end{tabular}   & \cellcolor[HTML]{9AFF99}0.9745                        & \cellcolor[HTML]{9AFF99}1                                      & \cellcolor[HTML]{9AFF99}0.0063                               & \cellcolor[HTML]{9AFF99}0.0063                            & \cellcolor[HTML]{9AFF99}0.4758                            \\ \cline{2-7} 
\multicolumn{1}{|c|}{}                                                                             & \begin{tabular}[c]{@{}c@{}}Medium\\ density\end{tabular} & \cellcolor[HTML]{9AFF99}0.7233                        & \cellcolor[HTML]{9AFF99}1                                      & \cellcolor[HTML]{9AFF99}0.9478                               & \cellcolor[HTML]{9AFF99}0.8006                            & \cellcolor[HTML]{FFCCC9}0.03939                           \\ \cline{2-7} 
\multicolumn{1}{|c|}{\multirow{-3}{*}{\begin{tabular}[c]{@{}c@{}}Small\\ Spheroids\end{tabular}}}  & \begin{tabular}[c]{@{}c@{}}Low\\ density\end{tabular}    & \cellcolor[HTML]{FFCCC9}0.0001                        & \cellcolor[HTML]{9AFF99}0.6915                                 & \cellcolor[HTML]{9AFF99}0.0059                               & \cellcolor[HTML]{9AFF99}0.2802                            & \cellcolor[HTML]{FFCCC9}0.0005                            \\ \hline
\multicolumn{1}{|c|}{}                                                                             & \begin{tabular}[c]{@{}c@{}}High\\ density\end{tabular}   & \cellcolor[HTML]{9AFF99}0.1728                        & \cellcolor[HTML]{9AFF99}{\color[HTML]{333333} 1}               & \cellcolor[HTML]{9AFF99}0.3677                               & \cellcolor[HTML]{9AFF99}0.3677                            & \cellcolor[HTML]{9AFF99}0.7242                            \\ \cline{2-7} 
\multicolumn{1}{|c|}{}                                                                             & \begin{tabular}[c]{@{}c@{}}Medium\\ density\end{tabular} & \cellcolor[HTML]{9AFF99}0.0881                        & \cellcolor[HTML]{9AFF99}{\color[HTML]{333333} 1}               & \cellcolor[HTML]{FFCCC9}0.0111                               & \cellcolor[HTML]{FFCCC9}0.0111                            & \cellcolor[HTML]{FFCCC9}0.0236                            \\ \cline{2-7} 
\multicolumn{1}{|c|}{\multirow{-3}{*}{\begin{tabular}[c]{@{}c@{}}Medium\\ Spheroids\end{tabular}}} & \begin{tabular}[c]{@{}c@{}}Low\\ density\end{tabular}    & \cellcolor[HTML]{9AFF99}0.0514                        & \cellcolor[HTML]{9AFF99}0.5948                                 & \cellcolor[HTML]{9AFF99}0.5008                               & \cellcolor[HTML]{9AFF99}0.0634                            & \cellcolor[HTML]{9AFF99}0.3408                            \\ \hline
\multicolumn{1}{|c|}{}                                                                             & \begin{tabular}[c]{@{}c@{}}High\\ density\end{tabular}   & \cellcolor[HTML]{9AFF99}0.6894                        & \cellcolor[HTML]{9AFF99}1                                      & \cellcolor[HTML]{9AFF99}0.0863                               & \cellcolor[HTML]{9AFF99}0.0863                            & \cellcolor[HTML]{9AFF99}0.1670                            \\ \cline{2-7} 
\multicolumn{1}{|c|}{}                                                                             & \begin{tabular}[c]{@{}c@{}}Medium\\ density\end{tabular} & \cellcolor[HTML]{FFCCC9}0.000                         & \cellcolor[HTML]{FFCCC9}0                                      & \cellcolor[HTML]{FFCCC9}0.0003                               & \cellcolor[HTML]{9AFF99}0.6035                            & \cellcolor[HTML]{9AFF99}0.2147                            \\ \cline{2-7} 
\multicolumn{1}{|c|}{\multirow{-3}{*}{\begin{tabular}[c]{@{}c@{}}Big\\ Spheroids\end{tabular}}}    & \begin{tabular}[c]{@{}c@{}}Low\\ density\end{tabular}    & \cellcolor[HTML]{9AFF99}{\color[HTML]{333333} 0.1162} & \cellcolor[HTML]{9AFF99}1                                      & \cellcolor[HTML]{9AFF99}0.2348                               & \cellcolor[HTML]{9AFF99}0.6712                            & \cellcolor[HTML]{9AFF99}0.3775                            \\ \hline
\end{tabular}
\caption{This data shows the $p-values$ of the Shapiro-Wilk test of normality for the computational simulations of Figure \ref{fig:comp_cel_lines}.  The green colored boxes  follow normal distributions. The red colored boxes do not. }
\label{table:normality_check_vel_red}
\end{table}

\begin{table}[h!]
\centering
\begin{tabular}{cc|c|c|c|c|c|}
\cline{3-7}
                                                                                                   &                                                          & Area                                             & \begin{tabular}[c]{@{}c@{}}Number of \\ spheroids\end{tabular}       & \begin{tabular}[c]{@{}c@{}}Cells per\\ spheroid\end{tabular} & \begin{tabular}[c]{@{}c@{}}Number\\ of cells\end{tabular} & \begin{tabular}[c]{@{}c@{}}Volume\\ per cell\end{tabular} \\ \hline
\multicolumn{1}{|c|}{}                                                                             & \begin{tabular}[c]{@{}c@{}}High\\ density\end{tabular}   & \cellcolor[HTML]{FFCCC9}                         & \cellcolor[HTML]{FFCCC9}                                             & \cellcolor[HTML]{9AFF99}                                     & \cellcolor[HTML]{9AFF99}                                  & \cellcolor[HTML]{FFCCC9}                                  \\ \cline{2-2}
\multicolumn{1}{|c|}{}                                                                             & \begin{tabular}[c]{@{}c@{}}Medium\\ density\end{tabular} & \cellcolor[HTML]{FFCCC9}                         & \cellcolor[HTML]{FFCCC9}                                             & \cellcolor[HTML]{9AFF99}                                     & \cellcolor[HTML]{9AFF99}                                  & \cellcolor[HTML]{FFCCC9}                                  \\ \cline{2-2}
\multicolumn{1}{|c|}{\multirow{-3}{*}{\begin{tabular}[c]{@{}c@{}}Small\\ Spheroids\end{tabular}}}  & \begin{tabular}[c]{@{}c@{}}Low\\ density\end{tabular}    & \multirow{-3}{*}{\cellcolor[HTML]{FFCCC9}0.0387} & \multirow{-3}{*}{\cellcolor[HTML]{FFCCC9}0.0157}                     & \multirow{-3}{*}{\cellcolor[HTML]{9AFF99}0.6980}             & \multirow{-3}{*}{\cellcolor[HTML]{9AFF99}0.0910}          & \multirow{-3}{*}{\cellcolor[HTML]{FFCCC9}0.0027}          \\ \hline
\multicolumn{1}{|c|}{}                                                                             & \begin{tabular}[c]{@{}c@{}}High\\ density\end{tabular}   & \cellcolor[HTML]{9AFF99}                         & \cellcolor[HTML]{FFCCC9}{\color[HTML]{333333} }                      & \cellcolor[HTML]{9AFF99}                                     & \cellcolor[HTML]{9AFF99}                                  & \cellcolor[HTML]{9AFF99}                                  \\ \cline{2-2}
\multicolumn{1}{|c|}{}                                                                             & \begin{tabular}[c]{@{}c@{}}Medium\\ density\end{tabular} & \cellcolor[HTML]{9AFF99}                         & \cellcolor[HTML]{FFCCC9}{\color[HTML]{333333} }                      & \cellcolor[HTML]{9AFF99}                                     & \cellcolor[HTML]{9AFF99}                                  & \cellcolor[HTML]{9AFF99}                                  \\ \cline{2-2}
\multicolumn{1}{|c|}{\multirow{-3}{*}{\begin{tabular}[c]{@{}c@{}}Medium\\ Spheroids\end{tabular}}} & \begin{tabular}[c]{@{}c@{}}Low\\ density\end{tabular}    & \multirow{-3}{*}{\cellcolor[HTML]{9AFF99}0.1164} & \multirow{-3}{*}{\cellcolor[HTML]{FFCCC9}{\color[HTML]{333333} 0.0}} & \multirow{-3}{*}{\cellcolor[HTML]{9AFF99}0.3461}             & \multirow{-3}{*}{\cellcolor[HTML]{9AFF99}0.1732}          & \multirow{-3}{*}{\cellcolor[HTML]{9AFF99}0.2786}          \\ \hline
\multicolumn{1}{|c|}{}                                                                             & \begin{tabular}[c]{@{}c@{}}High\\ density\end{tabular}   & \cellcolor[HTML]{9AFF99}                         & \cellcolor[HTML]{FFCCC9}                                             & \cellcolor[HTML]{9AFF99}                                     & \cellcolor[HTML]{9AFF99}                                  & \cellcolor[HTML]{FFCCC9}                                  \\ \cline{2-2}
\multicolumn{1}{|c|}{}                                                                             & \begin{tabular}[c]{@{}c@{}}Medium\\ density\end{tabular} & \cellcolor[HTML]{9AFF99}                         & \cellcolor[HTML]{FFCCC9}                                             & \cellcolor[HTML]{9AFF99}                                     & \cellcolor[HTML]{9AFF99}                                  & \cellcolor[HTML]{FFCCC9}                                  \\ \cline{2-2}
\multicolumn{1}{|c|}{\multirow{-3}{*}{\begin{tabular}[c]{@{}c@{}}Big\\ Spheroids\end{tabular}}}    & \begin{tabular}[c]{@{}c@{}}Low\\ density\end{tabular}    & \multirow{-3}{*}{\cellcolor[HTML]{9AFF99}0.3104} & \multirow{-3}{*}{\cellcolor[HTML]{FFCCC9}0.0}                        & \multirow{-3}{*}{\cellcolor[HTML]{9AFF99}0.1775}             & \multirow{-3}{*}{\cellcolor[HTML]{9AFF99}0.4721}          & \multirow{-3}{*}{\cellcolor[HTML]{FFCCC9}0.0030}          \\ \hline
\end{tabular}
\caption{This data shows the $p-values$ of the Levene test of variance normality for the computational simulations of Figure \ref{fig:comp_area} and \ref{fig:comp_vol_per_cell}. The variances from the data of the green colored boxes i homogeneous, meanwhile the variances from the data of the red colored boxes is not. }
\label{table:levene}
\end{table}

\begin{table}[h!]
\centering
\begin{tabular}{cc|c|c|c|c|c|}
\cline{3-7}
                                                                                                   &                                                          & Area                                             & \begin{tabular}[c]{@{}c@{}}Number of \\ spheroids\end{tabular}          & \begin{tabular}[c]{@{}c@{}}Cells per\\ spheroid\end{tabular} & \begin{tabular}[c]{@{}c@{}}Number\\ of cells\end{tabular} & \begin{tabular}[c]{@{}c@{}}Volume\\ per cell\end{tabular} \\ \hline
\multicolumn{1}{|c|}{}                                                                             & \begin{tabular}[c]{@{}c@{}}High\\ density\end{tabular}   & \cellcolor[HTML]{9AFF99}                         & \cellcolor[HTML]{FFCCC9}                                                & \cellcolor[HTML]{9AFF99}                                     & \cellcolor[HTML]{9AFF99}                                  & \cellcolor[HTML]{9AFF99}                                  \\ \cline{2-2}
\multicolumn{1}{|c|}{}                                                                             & \begin{tabular}[c]{@{}c@{}}Medium\\ density\end{tabular} & \cellcolor[HTML]{9AFF99}                         & \cellcolor[HTML]{FFCCC9}                                                & \cellcolor[HTML]{9AFF99}                                     & \cellcolor[HTML]{9AFF99}                                  & \cellcolor[HTML]{9AFF99}                                  \\ \cline{2-2}
\multicolumn{1}{|c|}{\multirow{-3}{*}{\begin{tabular}[c]{@{}c@{}}Small\\ Spheroids\end{tabular}}}  & \begin{tabular}[c]{@{}c@{}}Low\\ density\end{tabular}    & \multirow{-3}{*}{\cellcolor[HTML]{9AFF99}0.1401} & \multirow{-3}{*}{\cellcolor[HTML]{FFCCC9}0}                             & \multirow{-3}{*}{\cellcolor[HTML]{9AFF99}0.0526}             & \multirow{-3}{*}{\cellcolor[HTML]{9AFF99}0.3147}          & \multirow{-3}{*}{\cellcolor[HTML]{9AFF99}0.1514}          \\ \hline
\multicolumn{1}{|c|}{}                                                                             & \begin{tabular}[c]{@{}c@{}}High\\ density\end{tabular}   & \cellcolor[HTML]{9AFF99}                         & \cellcolor[HTML]{FFCCC9}{\color[HTML]{333333} }                         & \cellcolor[HTML]{9AFF99}                                     & \cellcolor[HTML]{9AFF99}                                  & \cellcolor[HTML]{9AFF99}                                  \\ \cline{2-2}
\multicolumn{1}{|c|}{}                                                                             & \begin{tabular}[c]{@{}c@{}}Medium\\ density\end{tabular} & \cellcolor[HTML]{9AFF99}                         & \cellcolor[HTML]{FFCCC9}{\color[HTML]{333333} }                         & \cellcolor[HTML]{9AFF99}                                     & \cellcolor[HTML]{9AFF99}                                  & \cellcolor[HTML]{9AFF99}                                  \\ \cline{2-2}
\multicolumn{1}{|c|}{\multirow{-3}{*}{\begin{tabular}[c]{@{}c@{}}Medium\\ Spheroids\end{tabular}}} & \begin{tabular}[c]{@{}c@{}}Low\\ density\end{tabular}    & \multirow{-3}{*}{\cellcolor[HTML]{9AFF99}0.4268} & \multirow{-3}{*}{\cellcolor[HTML]{FFCCC9}{\color[HTML]{333333} 0.0012}} & \multirow{-3}{*}{\cellcolor[HTML]{9AFF99}0.7122}             & \multirow{-3}{*}{\cellcolor[HTML]{9AFF99}0.9616}          & \multirow{-3}{*}{\cellcolor[HTML]{9AFF99}0.1146}          \\ \hline
\multicolumn{1}{|c|}{}                                                                             & \begin{tabular}[c]{@{}c@{}}High\\ density\end{tabular}   & \cellcolor[HTML]{9AFF99}                         & \cellcolor[HTML]{FFCCC9}                                                & \cellcolor[HTML]{9AFF99}                                     & \cellcolor[HTML]{9AFF99}                                  & \cellcolor[HTML]{FFCCC9}                                  \\ \cline{2-2}
\multicolumn{1}{|c|}{}                                                                             & \begin{tabular}[c]{@{}c@{}}Medium\\ density\end{tabular} & \cellcolor[HTML]{9AFF99}                         & \cellcolor[HTML]{FFCCC9}                                                & \cellcolor[HTML]{9AFF99}                                     & \cellcolor[HTML]{9AFF99}                                  & \cellcolor[HTML]{FFCCC9}                                  \\ \cline{2-2}
\multicolumn{1}{|c|}{\multirow{-3}{*}{\begin{tabular}[c]{@{}c@{}}Big\\ Spheroids\end{tabular}}}    & \begin{tabular}[c]{@{}c@{}}Low\\ density\end{tabular}    & \multirow{-3}{*}{\cellcolor[HTML]{9AFF99}0.1199} & \multirow{-3}{*}{\cellcolor[HTML]{FFCCC9}0.0002}                        & \multirow{-3}{*}{\cellcolor[HTML]{9AFF99}0.2876}             & \multirow{-3}{*}{\cellcolor[HTML]{9AFF99}0.6645}          & \multirow{-3}{*}{\cellcolor[HTML]{FFCCC9}0.0073}          \\ \hline
\end{tabular}
\caption{This data shows the $p-values$ of the Levene test of variance normality for the computational simulations of Figure \ref{fig:comp_cel_lines}. The variances from the data of the green colored boxes i homogeneous, meanwhile the variances from the data of the red colored boxes is not.}
\label{table:levene_vel_red}
\end{table}
\newpage
\begin{table}[h!]
\centering
\begin{tabular}{cc|c|c|c|c|c|}
\cline{3-7}
                                                                                                   &                                                          & Area                                                     & \begin{tabular}[c]{@{}c@{}}Number of \\ spheroids\end{tabular}                  & \begin{tabular}[c]{@{}c@{}}Cells per\\ spheroid\end{tabular} & \begin{tabular}[c]{@{}c@{}}Number\\ of cells\end{tabular} & \begin{tabular}[c]{@{}c@{}}Volume\\ per cell\end{tabular} \\ \hline
\multicolumn{1}{|c|}{}                                                                             & \begin{tabular}[c]{@{}c@{}}High\\ density\end{tabular}   & \cellcolor[HTML]{FFCCC9}                                 & \cellcolor[HTML]{FFCCC9}                                                        & \cellcolor[HTML]{9AFF99}                                     & \cellcolor[HTML]{9AFF99}                                  & \cellcolor[HTML]{FFCCC9}                                  \\ \cline{2-2}
\multicolumn{1}{|c|}{}                                                                             & \begin{tabular}[c]{@{}c@{}}Medium\\ density\end{tabular} & \cellcolor[HTML]{FFCCC9}                                 & \cellcolor[HTML]{FFCCC9}                                                        & \cellcolor[HTML]{9AFF99}                                     & \cellcolor[HTML]{9AFF99}                                  & \cellcolor[HTML]{FFCCC9}                                  \\ \cline{2-2}
\multicolumn{1}{|c|}{\multirow{-3}{*}{\begin{tabular}[c]{@{}c@{}}Small\\ Spheroids\end{tabular}}}  & \begin{tabular}[c]{@{}c@{}}Low\\ density\end{tabular}    & \multirow{-3}{*}{\cellcolor[HTML]{FFCCC9}Kruskal-Wallis} & \multirow{-3}{*}{\cellcolor[HTML]{FFCCC9}Kruskal-Wallis}                        & \multirow{-3}{*}{\cellcolor[HTML]{9AFF99}ANOVA}              & \multirow{-3}{*}{\cellcolor[HTML]{9AFF99}ANOVA}           & \multirow{-3}{*}{\cellcolor[HTML]{FFCCC9}Kruskal-Wallis}  \\ \hline
\multicolumn{1}{|c|}{}                                                                             & \begin{tabular}[c]{@{}c@{}}High\\ density\end{tabular}   & \cellcolor[HTML]{9AFF99}                                 & \cellcolor[HTML]{FFCCC9}{\color[HTML]{333333} }                                 & \cellcolor[HTML]{9AFF99}                                     & \cellcolor[HTML]{FFCCC9}                                  & \cellcolor[HTML]{9AFF99}                                  \\ \cline{2-2}
\multicolumn{1}{|c|}{}                                                                             & \begin{tabular}[c]{@{}c@{}}Medium\\ density\end{tabular} & \cellcolor[HTML]{9AFF99}                                 & \cellcolor[HTML]{FFCCC9}{\color[HTML]{333333} }                                 & \cellcolor[HTML]{9AFF99}                                     & \cellcolor[HTML]{FFCCC9}                                  & \cellcolor[HTML]{9AFF99}                                  \\ \cline{2-2}
\multicolumn{1}{|c|}{\multirow{-3}{*}{\begin{tabular}[c]{@{}c@{}}Medium\\ Spheroids\end{tabular}}} & \begin{tabular}[c]{@{}c@{}}Low\\ density\end{tabular}    & \multirow{-3}{*}{\cellcolor[HTML]{9AFF99}ANOVA}          & \multirow{-3}{*}{\cellcolor[HTML]{FFCCC9}{\color[HTML]{333333} Kruskal-Wallis}} & \multirow{-3}{*}{\cellcolor[HTML]{9AFF99}ANOVA}              & \multirow{-3}{*}{\cellcolor[HTML]{FFCCC9}Kruskal-Wallis}  & \multirow{-3}{*}{\cellcolor[HTML]{9AFF99}ANOVA}           \\ \hline
\multicolumn{1}{|c|}{}                                                                             & \begin{tabular}[c]{@{}c@{}}High\\ density\end{tabular}   & \cellcolor[HTML]{FFCCC9}                                 & \cellcolor[HTML]{FFCCC9}                                                        & \cellcolor[HTML]{9AFF99}                                     & \cellcolor[HTML]{9AFF99}                                  & \cellcolor[HTML]{FFCCC9}                                  \\ \cline{2-2}
\multicolumn{1}{|c|}{}                                                                             & \begin{tabular}[c]{@{}c@{}}Medium\\ density\end{tabular} & \cellcolor[HTML]{FFCCC9}                                 & \cellcolor[HTML]{FFCCC9}                                                        & \cellcolor[HTML]{9AFF99}                                     & \cellcolor[HTML]{9AFF99}                                  & \cellcolor[HTML]{FFCCC9}                                  \\ \cline{2-2}
\multicolumn{1}{|c|}{\multirow{-3}{*}{\begin{tabular}[c]{@{}c@{}}Big\\ Spheroids\end{tabular}}}    & \begin{tabular}[c]{@{}c@{}}Low\\ density\end{tabular}    & \multirow{-3}{*}{\cellcolor[HTML]{FFCCC9}Kruskal-Wallis} & \multirow{-3}{*}{\cellcolor[HTML]{FFCCC9}Kruskal-Wallis}                        & \multirow{-3}{*}{\cellcolor[HTML]{9AFF99}ANOVA}              & \multirow{-3}{*}{\cellcolor[HTML]{9AFF99}ANOVA}           & \multirow{-3}{*}{\cellcolor[HTML]{FFCCC9}Kruskal-Wallis}  \\ \hline
\end{tabular}
\caption{This table sum up which of the data groups presented in Figure \ref{fig:comp_area} and \ref{fig:comp_vol_per_cell} follow normal distributions, have homogeneus variances and therefore can be analysed through an ANOVA test(green) test and which ones do not and have to be analysed through a Kruskal-Wallis test(red).   }
\label{table:kruskal_anova}
\end{table}

\begin{table}[h!]
\centering
\begin{tabular}{cc|c|c|c|c|c|}
\cline{3-7}
                                                                                                   &                                                          & Area                                                     & \begin{tabular}[c]{@{}c@{}}Number of \\ spheroids\end{tabular}                  & \begin{tabular}[c]{@{}c@{}}Cells per\\ spheroid\end{tabular} & \begin{tabular}[c]{@{}c@{}}Number\\ of cells\end{tabular} & \begin{tabular}[c]{@{}c@{}}Volume\\ per cell\end{tabular} \\ \hline
\multicolumn{1}{|c|}{}                                                                             & \begin{tabular}[c]{@{}c@{}}High\\ density\end{tabular}   & \cellcolor[HTML]{FFCCC9}                                 & \cellcolor[HTML]{FFCCC9}                                                        & \cellcolor[HTML]{9AFF99}                                     & \cellcolor[HTML]{9AFF99}                                  & \cellcolor[HTML]{FFCCC9}                                  \\ \cline{2-2}
\multicolumn{1}{|c|}{}                                                                             & \begin{tabular}[c]{@{}c@{}}Medium\\ density\end{tabular} & \cellcolor[HTML]{FFCCC9}                                 & \cellcolor[HTML]{FFCCC9}                                                        & \cellcolor[HTML]{9AFF99}                                     & \cellcolor[HTML]{9AFF99}                                  & \cellcolor[HTML]{FFCCC9}                                  \\ \cline{2-2}
\multicolumn{1}{|c|}{\multirow{-3}{*}{\begin{tabular}[c]{@{}c@{}}Small\\ Spheroids\end{tabular}}}  & \begin{tabular}[c]{@{}c@{}}Low\\ density\end{tabular}    & \multirow{-3}{*}{\cellcolor[HTML]{FFCCC9}Kruskal-Wallis} & \multirow{-3}{*}{\cellcolor[HTML]{FFCCC9}Kuskal-Wallis}                         & \multirow{-3}{*}{\cellcolor[HTML]{9AFF99}ANOVA}              & \multirow{-3}{*}{\cellcolor[HTML]{9AFF99}ANOVA}           & \multirow{-3}{*}{\cellcolor[HTML]{FFCCC9}Kruskal-Wallis}  \\ \hline
\multicolumn{1}{|c|}{}                                                                             & \begin{tabular}[c]{@{}c@{}}High\\ density\end{tabular}   & \cellcolor[HTML]{9AFF99}                                 & \cellcolor[HTML]{FFCCC9}{\color[HTML]{333333} }                                 & \cellcolor[HTML]{FFCCC9}                                     & \cellcolor[HTML]{FFCCC9}                                  & \cellcolor[HTML]{FFCCC9}                                  \\ \cline{2-2}
\multicolumn{1}{|c|}{}                                                                             & \begin{tabular}[c]{@{}c@{}}Medium\\ density\end{tabular} & \cellcolor[HTML]{9AFF99}                                 & \cellcolor[HTML]{FFCCC9}{\color[HTML]{333333} }                                 & \cellcolor[HTML]{FFCCC9}                                     & \cellcolor[HTML]{FFCCC9}                                  & \cellcolor[HTML]{FFCCC9}                                  \\ \cline{2-2}
\multicolumn{1}{|c|}{\multirow{-3}{*}{\begin{tabular}[c]{@{}c@{}}Medium\\ Spheroids\end{tabular}}} & \begin{tabular}[c]{@{}c@{}}Low\\ density\end{tabular}    & \multirow{-3}{*}{\cellcolor[HTML]{9AFF99}ANOVA}          & \multirow{-3}{*}{\cellcolor[HTML]{FFCCC9}{\color[HTML]{333333} Kruskal-Wallis}} & \multirow{-3}{*}{\cellcolor[HTML]{FFCCC9}Kruskal-Wallis}     & \multirow{-3}{*}{\cellcolor[HTML]{FFCCC9}Kruskal-Wallis}  & \multirow{-3}{*}{\cellcolor[HTML]{FFCCC9}Kruskal-Wallis}  \\ \hline
\multicolumn{1}{|c|}{}                                                                             & \begin{tabular}[c]{@{}c@{}}High\\ density\end{tabular}   & \cellcolor[HTML]{FFCCC9}                                 & \cellcolor[HTML]{FFCCC9}                                                        & \cellcolor[HTML]{FFCCC9}                                     & \cellcolor[HTML]{9AFF99}                                  & \cellcolor[HTML]{FFCCC9}                                  \\ \cline{2-2}
\multicolumn{1}{|c|}{}                                                                             & \begin{tabular}[c]{@{}c@{}}Medium\\ density\end{tabular} & \cellcolor[HTML]{FFCCC9}                                 & \cellcolor[HTML]{FFCCC9}                                                        & \cellcolor[HTML]{FFCCC9}                                     & \cellcolor[HTML]{9AFF99}                                  & \cellcolor[HTML]{FFCCC9}                                  \\ \cline{2-2}
\multicolumn{1}{|c|}{\multirow{-3}{*}{\begin{tabular}[c]{@{}c@{}}Big\\ Spheroids\end{tabular}}}    & \begin{tabular}[c]{@{}c@{}}Low\\ density\end{tabular}    & \multirow{-3}{*}{\cellcolor[HTML]{FFCCC9}Kruskal-Wallis} & \multirow{-3}{*}{\cellcolor[HTML]{FFCCC9}Kruskal-Wallis}                        & \multirow{-3}{*}{\cellcolor[HTML]{FFCCC9}Kruskall-Wallis}    & \multirow{-3}{*}{\cellcolor[HTML]{9AFF99}ANOVA}           & \multirow{-3}{*}{\cellcolor[HTML]{FFCCC9}Kruskal-Wallis}  \\ \hline
\end{tabular}
\caption{This table sum up which of the data groups presented in Figure \ref{fig:comp_cel_lines}  follow normal distributions, have homogeneus variances and therefore can be analysed through an ANOVA test(green) test and which ones do not and have to be analysed through a Kruskal-Wallis test(red).   }
\label{table:kruskal_anova_red}
\end{table}

\end{document}